\documentclass{emulateapj}
\usepackage[intlimits,fleqn]{amsmath}

\newcommand{\D}[1]{\, d #1 \,}
\newcommand{\vc}[1]{\vec{#1}}
\newcommand{\avg}[1]{\left< #1 \right>}
\newcommand{\ft}[1]{\widetilde{#1}}

\newcommand{\eps}{\varepsilon}

\newcommand{\sourcedir}{\texttt{www.astro.wisc.edu$/\!\!\sim$lazarian$/$simulations}}

\newcommand{\inctab}[1]{\includegraphics[scale=.25]{#1}}

\shorttitle{Turbulence Spectra from Spectral Lines}
\shortauthors{Chepurnov \& Lazarian}

\begin{document}


\title{Turbulence Spectra from Spectral Lines: Tests of the Velocity Channel Analysis and Velocity Coordinate Spectrum Techniques}

\author{A. Chepurnov, A. Lazarian}
\affil{Department of Astronomy, University of Wisconsin, Madison, USA}

\begin{abstract}
Turbulence is a key element of the dynamics of astrophysical fluids, including those of interstellar medium, clusters of galaxies and circumstellar regions. Turbulent motions induce Doppler shifts of observable emission and absorption lines and this motivates studies of turbulence using precision spectroscopy. We provide high resolution numerical testing of the two promising techniques, namely, Velocity Channel Analysis and Velocity Coordinate Spectrum. We obtain an expression for the shot noise that the discretization of the numerical data entails and successfully test it. We show that numerical resolution required for recovering the underlying turbulent spectrum from observations depend on the spectral index of velocity fluctuations. Thus the low resolution testing may be misleading.
\end{abstract} 

\section{Introduction}

As a rule astrophysical fluids are turbulent and the turbulence is magnetized. This ubiquitous turbulence determines the transport properties of interstellar medium (see \citealt{EF96}, \citealt{Stu01}, \citealt{Bal06}) and intracluster medium (see \citealt{Sun03}, \citealt{Ens06}, \citealt{Laz06b}), many properties of Solar and stellar winds (see \citealt{HMG80}) etc. One may say that to understand heat conduction, propagation of cosmic rays and electromagnetic radiation in different astrophysical environments it is absolutely essential to understand the properties of underlying magnetized turbulence. The fascinating processes of star formation (see \citealt{MKT02}, \citealt{Elm02}, \citealt{MLK04}) and interstellar chemistry (see \cite{Fal06} and references therein) are also intimately related to properties of magnetized compressible turbulence (see reviews by \citealt{ES04}). 

We should stress that while the density fluctuations are an indirect way of testing turbulence, the most valuable information is given by the velocity field statistics, encoded in spectrometric observations. To recover it we need to account for mapping of a velocity field in real space to position-position-velocity (PPV) observational data cubes. 

We have several options here. For subsonic or slow supersonic turbulence we can use techniques of velocity centroids, which allows us recovering of velocity spectrum (\citealt{EL05}). As it was shown there, this technique does not work for turbulence with high Mach number. For such turbulence Velocity Channel Analysis (VCA) (see \citealt{LP00}), which deals with 2-d slices of PPV, and Velocity Coordinate Spectrum (VCS) (see \citealt{LP00}, \citealt{LP06}, \citealt{LC06}) should be used.

VCA has been already used with both atomic hydrogen and molecular line data to recover the underlying turbulence spectra (see review by \citealt{Laz06c}). The technique has been tested with numerics for optically thin case in \cite{Laz01} and \cite{Esq03} and also with absorption effects in \cite{Pad06}. However, rather strong shot-noise was observed in the synthetic channel maps. For instance, numerical simulations for VCA in \cite{Esq03} show that discretization over line of sight may distort substantially the spectrum at high wavenumbers. It was also shown, that its impact can be minimized, when velocity slice is made thickest possible for the thin (velocity-dominated) regime. The shot-noise was also observed to decrease with the increase of the cube size used for analysis. Nevertheless, the exact origin of it was somewhat unclear. This effect seemed to confuse other researchers who started to wonder about the practical utility of the technique with the real spectroscopic data \citep{Miv03}.

The VCS is a more recent development. Although the formulae for it were obtained in \cite{LP00}, the actual application of it has just started \citep{Che07}. No extensive testing of the technique has been performed before this work, as far as we know.

In this paper we deal with the numerical testing for both VCA and VCS techniques. In particular, we subject the origin of shot noise to scrutiny in this paper.

In what follows we provide some facts about the VCA and VCS in Sect. \ref{sect:bf}, analyze shot-noise in Sect. \ref{sect:sn}, describe the results of numerical testing in Sect. \ref{sect:res} and provide the summary in Sect. \ref{sect:sum}.

\section{Basic facts} \label{sect:bf}

\begin{figure}[p] 
\begin{center}
\plotone{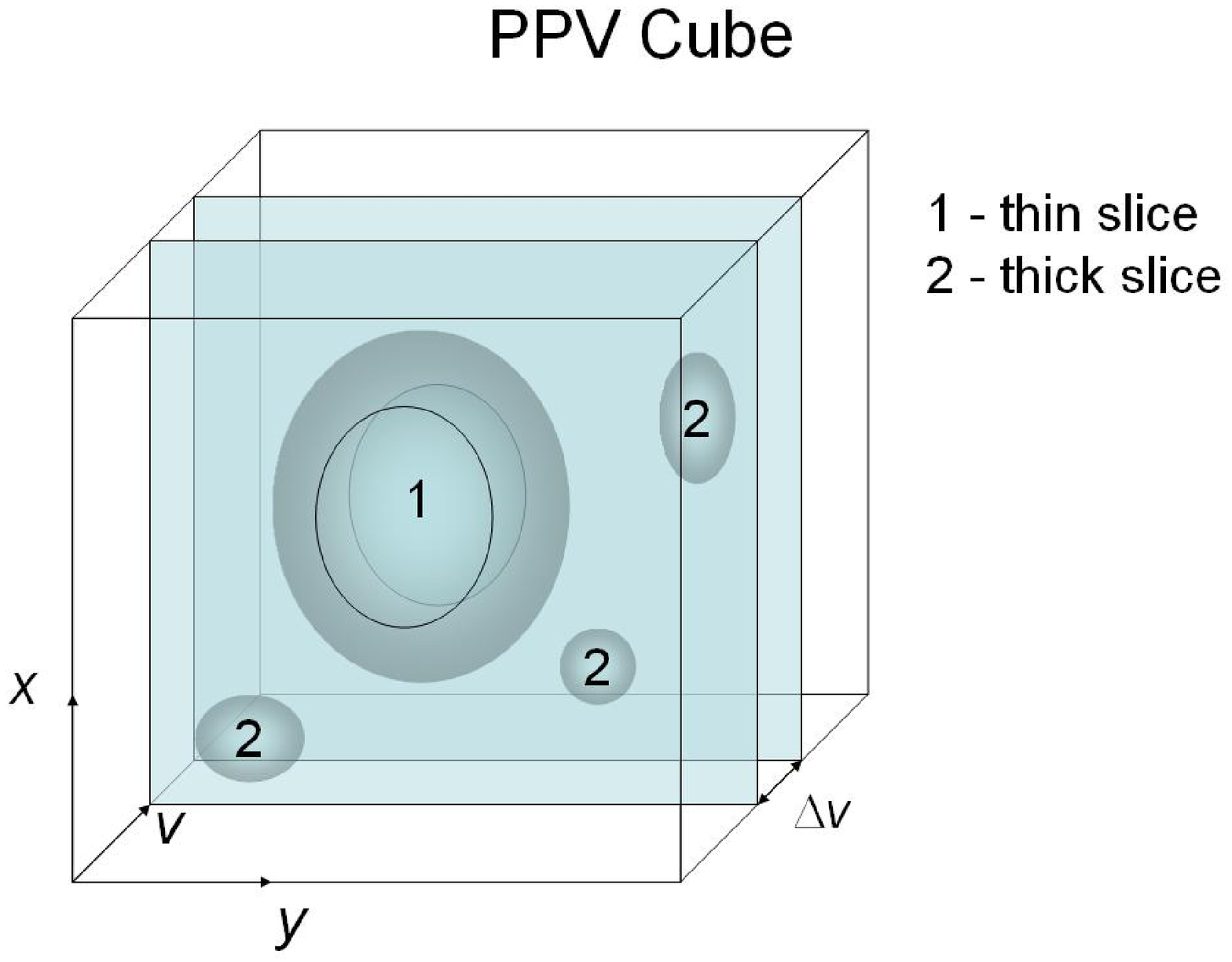}
\end{center}
\caption{Illustration to VCA channel modes. Eddies, exceeding channel width are velocity-dominated, those with extent over $v$ less than $\Delta v$ are density-dominated. Velocity-dominated spectrum has the slope $(9-\alpha_v)/2$ (for steep density spectrum), the density-dominated one has the slope equal to $\alpha_\eps$ ($\alpha_v$  and $\alpha_\eps$ are the slopes of velocity and density power spectra)\label{fig:VCA_res}} 
\end{figure}

We show in Fig. \ref{fig:VCA_res} the essence of the practical data handling with the VCA. The data in the PPV data cube is being analyzed in the slices of $\Delta V$ thickness. The images of the eddies can have an extend both larger and smaller than $\Delta V$. The former case is termed by \cite{LP00} ``thin slice'', the latter is termed ``thick slice''.

\begin{figure}[p] 
\begin{center}
\plotone{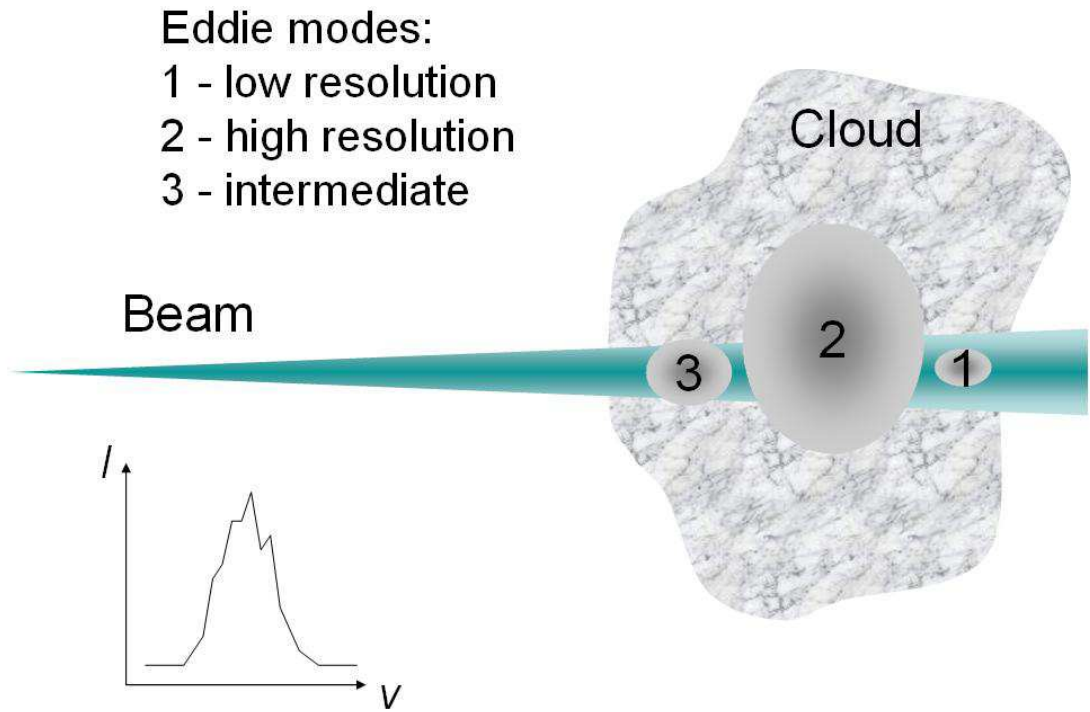}
\end{center}
\caption{Illustration to VCS resolution regimes. Eddies within beam size are in low resolution mode and cause steepening of $P_1$. The ones exceeding the beam size are in high-resolution mode and correspondent $P_1$ is shallower. Therefore high-resolution mode is more preferable regarding to the signal-to-noise ratio.\label{fig:VCS_res}} 
\end{figure}

In Fig. \ref{fig:VCS_res} we illustrate the essence of the VCS technique. Depending on the resolution one may or not resolve the spatial extent of the eddies under study.

Below we briefly overview the main analytical results, obtained in \cite{LP00}, \cite{LP04}, \cite{LP06} and \cite{LC06}, which are relevant for this paper.

Our calculations are based on the following expression for a spectral line signal, measured at velocity $v_0$ and given beam position: 
\begin{equation} \label{eq:sig_rat}
S(v_0)
  = \int w(\vc{r}) \D{\vc{r}} \eps(\vc{r}) f(v_r(\vc{r}) + v_r^{reg}(\vc{r}) - v_0),
\end{equation}
where $\eps$ is normalized emissivity, $f$ is convolution between channel sensitivity function and Maxwellian distribution, defined by temperature of emitting medium, and $w$ is a window function defined as follows: 
\begin{equation} \label{eq:w}
w(\vc{r})
  \equiv \frac{1}{r^2} w_b(\phi,\theta) w_\eps(\vc{r}),
\end{equation}
where $w_b$ is an instrument beam, $w_\eps$ is a window function defining the extend of the observed object.

If we take a Fourier transform of $S$ and correlate this quantity taken in two directions $1$ and $2$, we have the following measure, which can be used as a starting point for the mathematical formulation of both VCA and VCS (see \cite{LC06} for details): 
\begin{equation} \label{eq:K12} 
\begin{split}
K_{12}(k_v) 
  & \equiv \avg{\ft{S_1}(k_v) \ft{S_2}^*(k_v)} \\ 
  & =\ft{f}^2(k_v) \int w_{12}(\vc{r}) \D{\vc{r}} C_\eps(\vc{r}) \\
  & \cdot \exp \left(-\frac{k_v^2}{2} D_z(\vc{r}) - i k_v b z \right),
\end{split}
\end{equation}
where
\begin{equation} \label{eq:w12_def} 
w_{12}(\vc{r}) 
  \equiv \int w_1(\vc{r'}) w_2(\vc{r'}+\vc{r}) \D{\vc{r'}} 
\end{equation} 
and $D_z$ is a velocity structure tensor projection:
\begin{equation} \label{eq:Dz_def} 
D_z(\vc{r}-\vc{r'})
  \equiv \avg{(v_z(\vc{r})-v_z(\vc{r'}))^2}, 
\end{equation}

Expression [\ref{eq:K12}] is true provided that the following conditions are satisfied: 1. fields involved are homogeneous, 2. velocity is Gaussian, 3. absorption effects are negligible\footnote{In this case analysis is more simple if we use the Fourier space. Absorption effects can be described only in real space, which implies using of structure functions instead, see \cite{LP04}, \cite{LP06}}, 4. beam separation as well as beam width are small enough to neglect the difference between $v_r$ and $v_z$ (we consider $z$-axis to be a bisector of the angle between beams). We also assume that 5. $v_z^{reg}(\vc{r})$ depends only on $z$ and admits a linear approximation with angular coefficient $b$. The relative importance of these conditions and the way to avoid some of them within the VCA and the VCS framework are discussed in \cite{LP04} and \cite{LP06}. However, for the sake of simplicity, we do not discuss the most general case here.

If we set $w_{12}=w_{b,a}(\vc{R}-\vc{R}_{12})w_{\eps,a}(z)$ ($\vc{R}_{12}$ is a beam separation in picture plane, $w_{b,a}$ and $w_{\eps,a}$ are auto-convolutions of $w_b$ and $w_\eps$), which corresponds to the case of remote emitting object, and take Fourier transform over $\vc{R}_{12}$, we have a power spectrum, designated in VCA as $P_s$ (\citealt{LP00}): 
\begin{equation} \label{eq:Ps}
\begin{split}
P_s(\vc{K},k_v)
   &=\ft{f}^2(k_v) \ft{w_b}^2(\vc{K}) \int w_\eps(z) \D{\vc{r}} \\
   & \cdot e^{-i(\vc{K}\vc{R}+b k_v z)}
     C_\eps(\vc{r}) \exp \left(-\frac{k_v^2}{2} D_z(\vc{r}) \right).
\end{split}
\end{equation}

It is convenient to introduce $P_3$ as a 3D power spectrum of a \textit{PPV data cube} with assumption of infinite resolution over $v$ and $\vc{R}$, which, for instance, corresponds to the conditions for the unprocessed synthetic PPV cubes: 
\begin{equation} \label{eq:P3}
\begin{split}
P_3(\vc{K},k_v)
  & \equiv \avg{|\ft{S}(\vc{K},k_v)|^2}
  = \int w_\eps(z) \D{\vc{r}}  \\
  & \cdot e^{-i(\vc{K}\vc{R}+b k_v z)} C_\eps(\vc{r}) \exp \left(-\frac{k_v^2}{2} D_z(\vc{r}) \right).
\end{split}
\end{equation}
In this case $P_s(\vc{K},k_v) = \ft{f}^2(k_v) \ft{w_b}^2(\vc{K}) P_3(\vc{K},k_v)$.

The VCA predicts the behavior of the quantity
\begin{equation} \label{eq:P2}
\begin{split}
P_2(\vc{K})
  &\equiv \int_{-\infty}^\infty \D{k_v} P_s(\vc{K},k_v) \\
  &= \ft{w_b}^2(\vc{K}) \int_{-\infty}^\infty \D{k_v} \ft{f}^2(k_v) P_3(\vc{K},k_v),
\end{split}
\end{equation}
which can be easily determined from observations. For the channel which effective width\footnote{It is approximately equal to $\sqrt(\delta v_{ch}^2+(2 v_T)^2)$, where $\delta v_{ch}$ is channel width and $v_T$ is thermal velocity of emitting atoms.} is small enough (see Fig. \ref{fig:VCA_res}) $P_2$ is \textit{velocity dominated} regime with the slope slope $(9-\alpha_v)/2$, provided that the density has a steep spectrum, i.e. $\alpha_\eps>3$. the later can be established from column density measured with infinitely wide slices.  Then the  spectum is \textit{density dominated}, and the slope is $\alpha_\eps$. Here $\alpha_v$  and $\alpha_\eps$ are the slopes of velocity and density power spectra. For a more detailed analysis see \cite{LP00}, while the predictions for different variants of velocity-dominated regime are summarized in Tab. \ref{tab:P2_slopes}.

The one-dimensional spectrum $P_1$ is the subject of the VCS studies. It corresponds to the case when the two beams, involved in $K_{12}$, coincide, i.e. 
\begin{equation} \label{eq:P1_orig} 
\begin{split}
P_1(k_v)
  & \equiv K_{12}(k_v)\vert_{\vc{R}_{12}=0} 
  = \ft{f}^2(k_v) \int w_{11}(\vc{r}) \D{\vc{r}} \\
  &  \cdot C_\eps(\vc{r}) \exp \left(-\frac{k_v^2}{2} D_z(\vc{r}) - i k_v b z \right).
\end{split}
\end{equation}

In terms of $P_3$ the one-dimensional spectrum can be written as follows:
\begin{equation} \label{eq:P1}
P_1(k_v)
  = \ft{f}^2(k_v) \int \D{\vc{K}} \ft{w_b}^2(\vc{K}) P_3(\vc{K},k_v).
\end{equation}
As it was shown in \cite{LP06} and \cite{LC06}, here we also have two regimes of spectum. They depend on beamwidth, see Fig. \ref{fig:VCS_res} for illustration. For \textit{high resolution mode} ($k_v$ is less than velocity variance on the beam scale) the slope of $P_1$ is $2/(\alpha_v - 3)$, otherwise, in \textit{low resolution mode} it is $6/(\alpha_v - 3)$ (steep density spectrum is assumed for the both cases). More complete list of the predictions for $P_1$ is presented in Tab. \ref{tab:P1_slopes}.

If we compare Eqs. [\ref{eq:P2}] and [\ref{eq:P1}], we see that the VCA and the VCS provide complimentary approaches and are related to the same quantity, $P_3$. Their properties must be closely related therefore.

\section{Shot-noise} \label{sect:sn}

In \cite{Esq03} it was shown, that the quality of the VCA simulations strongly depends on the number of points along the line of sight. The underlying interference was termed \textit{shot-noise}  to reflect is stochastic nature: insufficient statistics for the signal in spectrometer channels. We revealed the same type of noise for VCS. In this case, $P_1$ saturates to a constant level, which decreases an available interval of power law (see Fig. \ref{fig:VCA_VCS_3_67_unfilt}, right). Below we address the condition needed to avoid it.

To have a realistic spectral line we need a velocity field to be resolved over line of sight well enough (see also \citealt{IS03}). This means, that the typical scale over $z$, for which velocity stays within a single channel near its extremum, should have enough discretization points, because such extrema give the dominant contribution to the spectrometer signal. 

Below we estimate the required resolution. To do so, we need to bind the correspondent typical scale over line of sight with the channel width. The natural choice is calculating velocity dispersion (which itself can be associated with the channel width) for the wavelengths shorter than this scale. As mentioned above, the contribution of the velocity profile shapes other than extrema, provided by the remaining harmonics, can be neglected. Therefore, the scale found by reverting of this expression is what we need.


In other words, we divide the 1-d velocity spectrum domain into the intervals separated by the scale in question $L_\Delta$, and, as spoken, assume that the harmonics in its low-wavenumber part provide an extremum of velocity. For the higher harmonics, in order to place the bulk of the signal inside one channel, we should demand, that the correspondent variance $\sigma(L_\Delta)$ is about half a channel width:
\begin{equation} \label{eq:sigmabalance}
\sigma(L_\Delta) 
  \approx \dfrac{\sigma(\infty)}{2 N_\sigma},
\end{equation} 
where $L_\Delta$ is a scale, at which velocity stays within a single channel near its extremum, $N_\sigma$ is a number of channels per velocity variance, and $\sigma(L)$ is a velocity variance for scales smaller than $L$, expressed through 1-d power spectrum\footnote{Should not be confused with $P_1$, that refers to the PPV space} $P_{1d}$: 
\begin{equation} \label{eq:sigma}
\sigma^2(L)
  = 2 \int_{\frac{2\pi}{L}}^\infty P_{1d}(k) \D{k}.
\end{equation} 

If a 3-d power spectrum has a cutoff at $k_0=2\pi/L$, where $L$ is an injection scale, $P_{1d}$ has the following form for the isotropic 3-d spectrum of $v_z$ 
\begin{equation} \label{eq:P_1d}
P_{1d}
  \sim \frac{1}{\alpha_v-2} \left\{ 
    \begin{array}{ll} 
      k^{-(\alpha_v-2)},   & |k| > k_0 \\
      k_0^{-(\alpha_v-2)}, & |k| \le k_0 \\
    \end{array} 
  \right.
\end{equation} 

Then, having solved Eq. [\ref{eq:sigmabalance}] for $L_\Delta$ we get
\begin{equation} \label{eq:L_delta}
L_\Delta 
  \approx \left(\frac{2 N_\sigma}{\sqrt{\alpha_v-2}}\right)^{-\frac{2}{\alpha_v-3}} L~.
\end{equation} 

If $n_\Delta$ is a number of points per $L_\Delta$, then the total number of points along $z$ is as follows: 
\begin{equation} \label{eq:Nz}
N_z
  \approx n_\Delta \frac{L_s}{L} \left(\frac{2 N_\sigma}{\sqrt{\alpha_v-2}}\right)^{\frac{2}{\alpha_v-3}} 
\end{equation} 
where $L_s$ is the size of emitting structure, $n_\Delta$ is determined from simulations and is experimentally found to be around 6.

The requirements for $N_z$ are rather demanding. If we have taken our typical settings: 128 channels per triple line width and $L_s/L=2$, then for $\alpha_v = 4$ $N_z$ there should be at least $1.1\cdot10^4$ points, for $\alpha_v = 11/3$ the corresponding number is $4.3\cdot10^5$, while for $\alpha_v = 3.5$ the number is as big as $1.7\cdot10^7$. Therefore it is not surprising that without this knowledge some researchers could get puzzled by their results (e.g. \citealt{Miv03}).

Condition [\ref{eq:Nz}] guarantees, that the spectrum is clear from shot-noise for all available $k_v$'s, i.e. up to $2\pi/\Delta v$, where $\Delta v$ is the channel width. This estimation refers to the case without smoothing along angular coordinates, i.e. for the high-resolution regime. Low-resolution VCS spectrum drops faster and therefore the requirements are even more restrictive.

\section{Results} \label{sect:res}

\begin{figure}[p]
\begin{center}
\begin{tabular}{lll}
\inctab{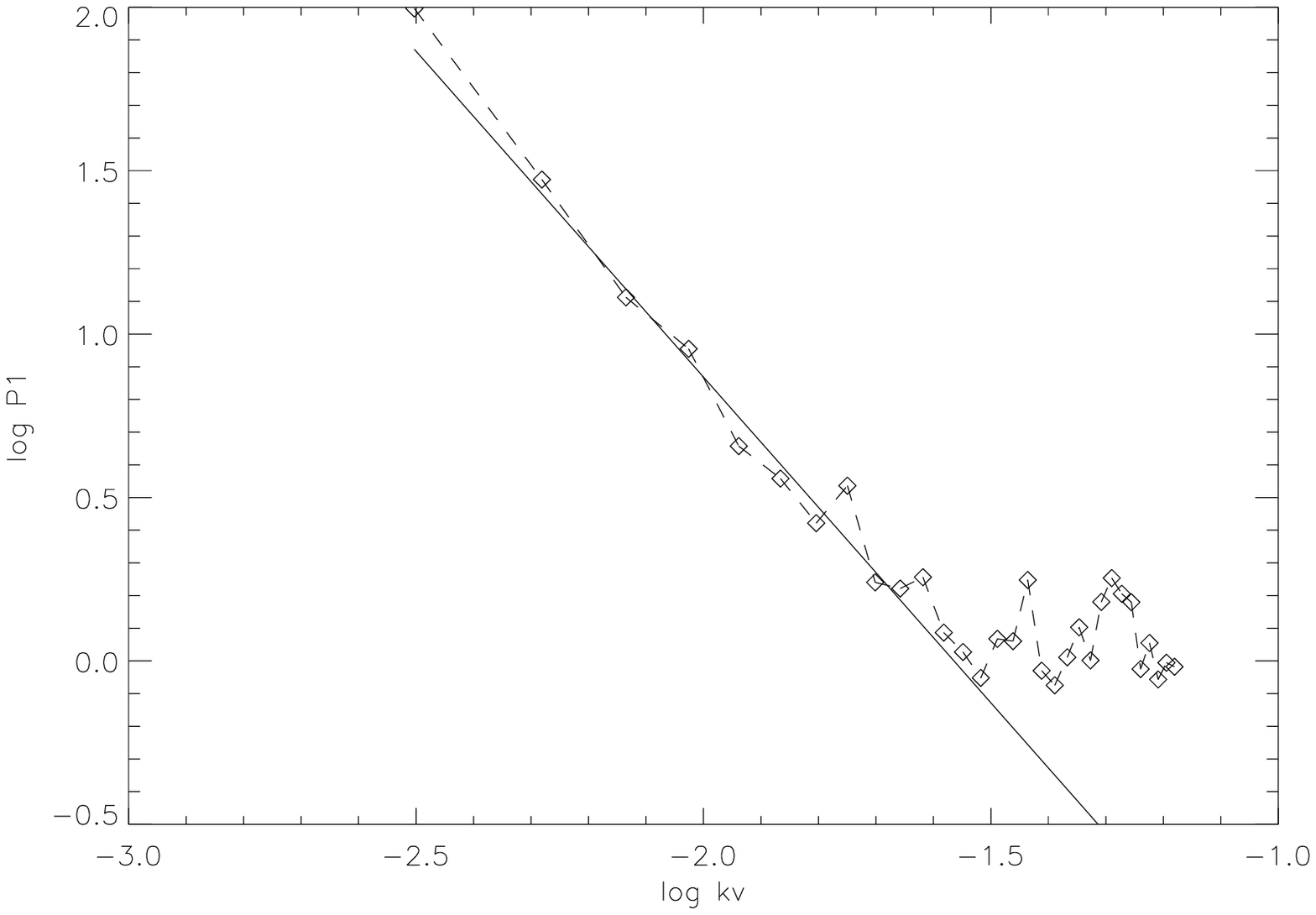} &
\inctab{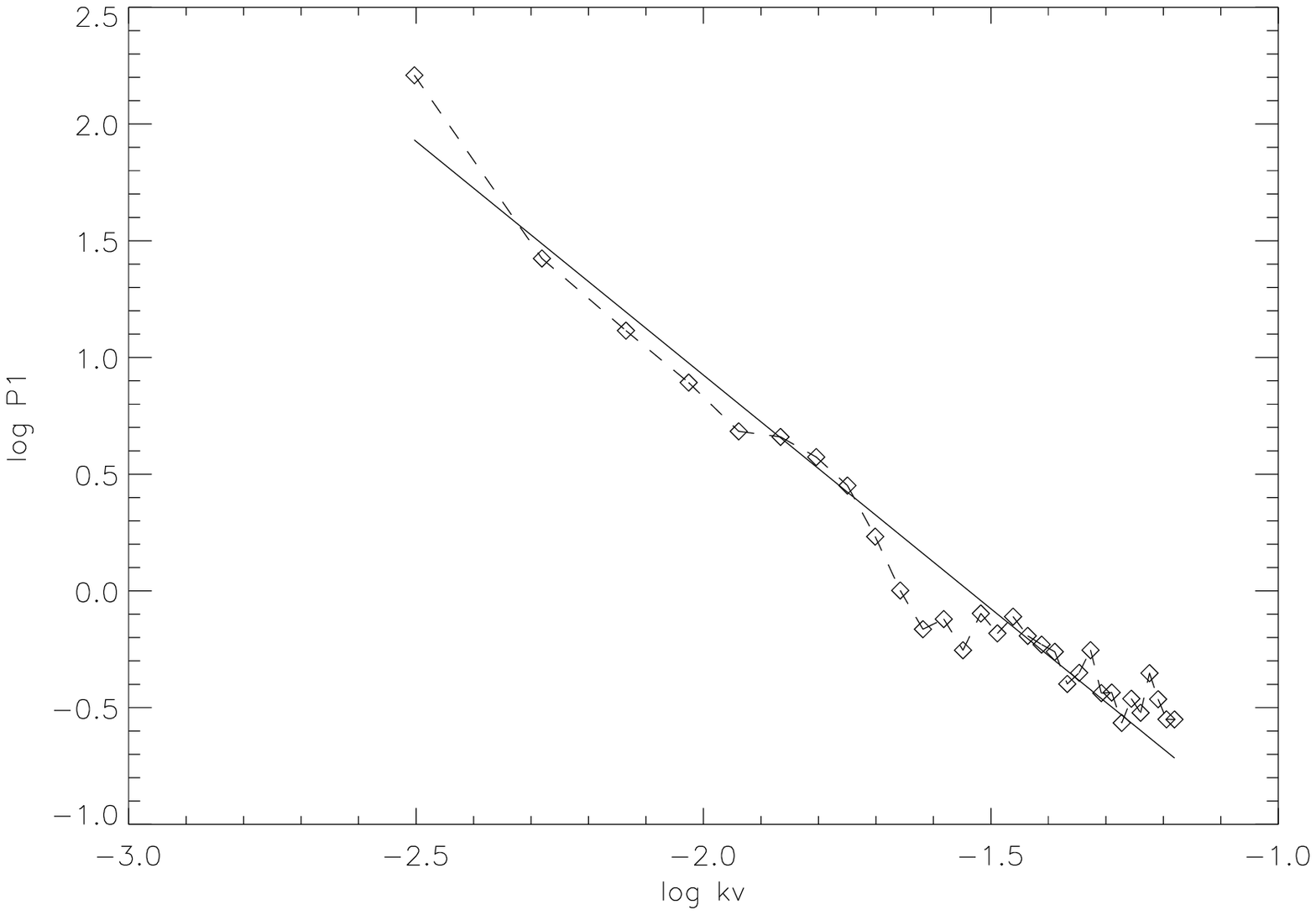} &
\inctab{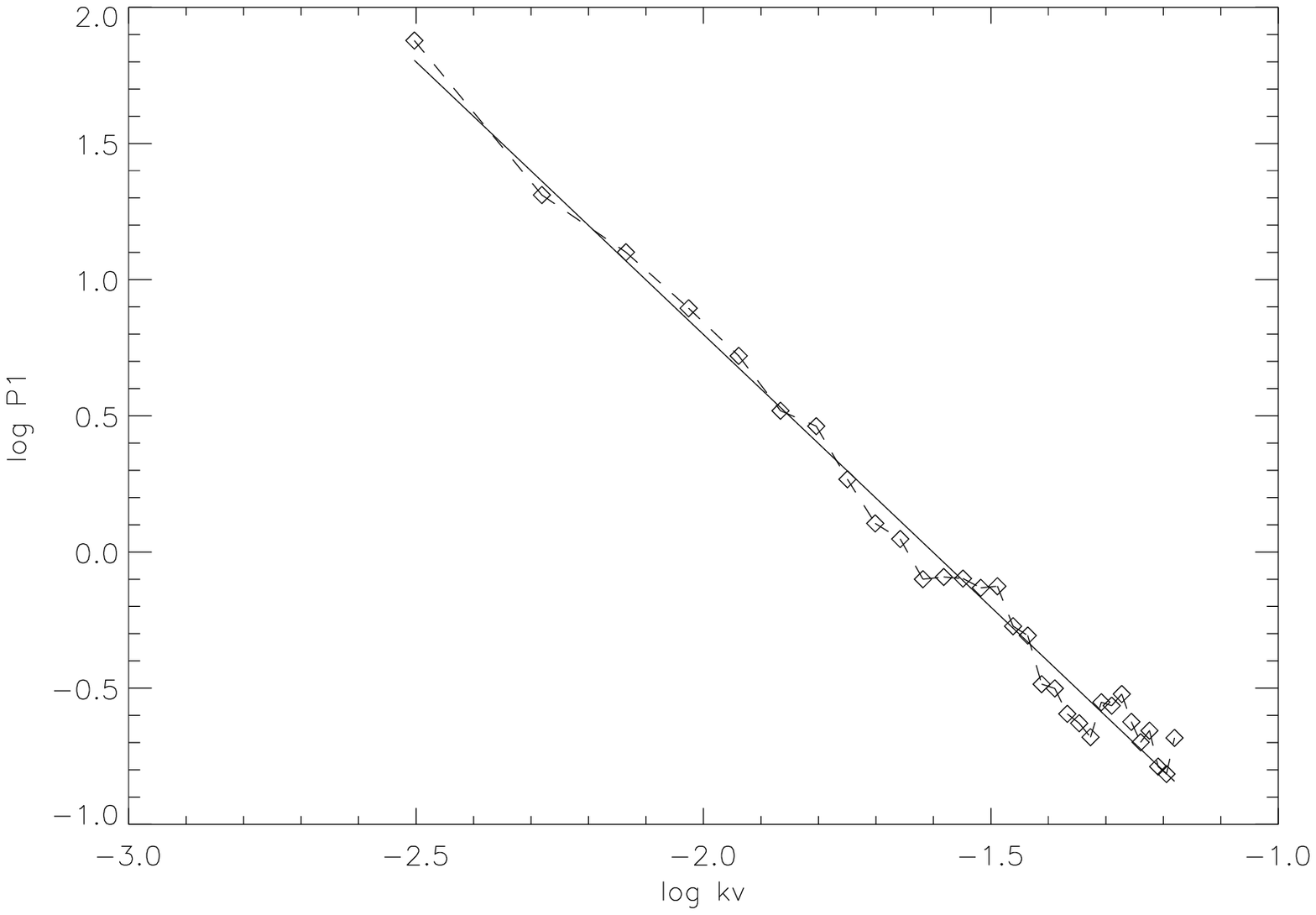} \\
\inctab{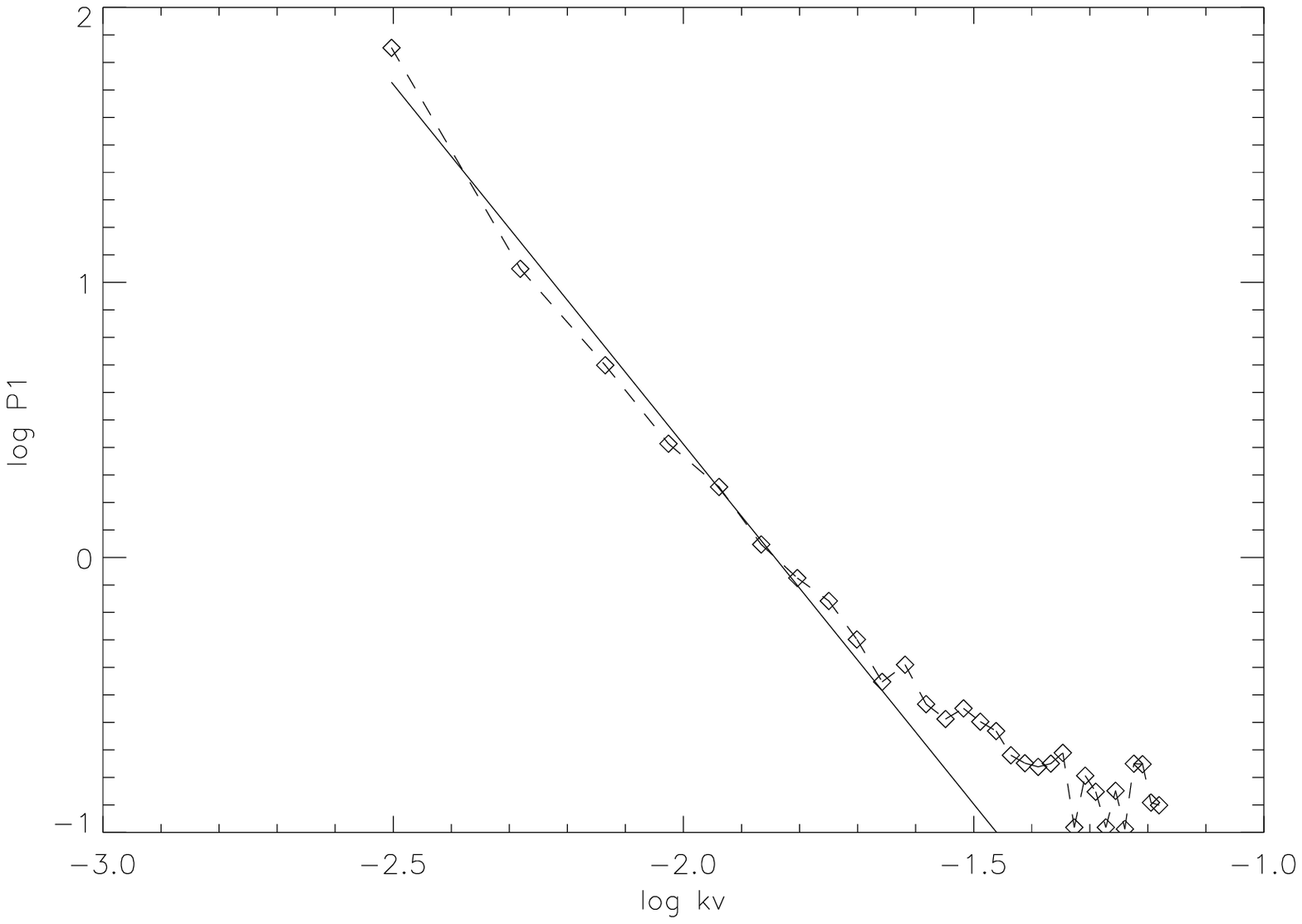} &
\inctab{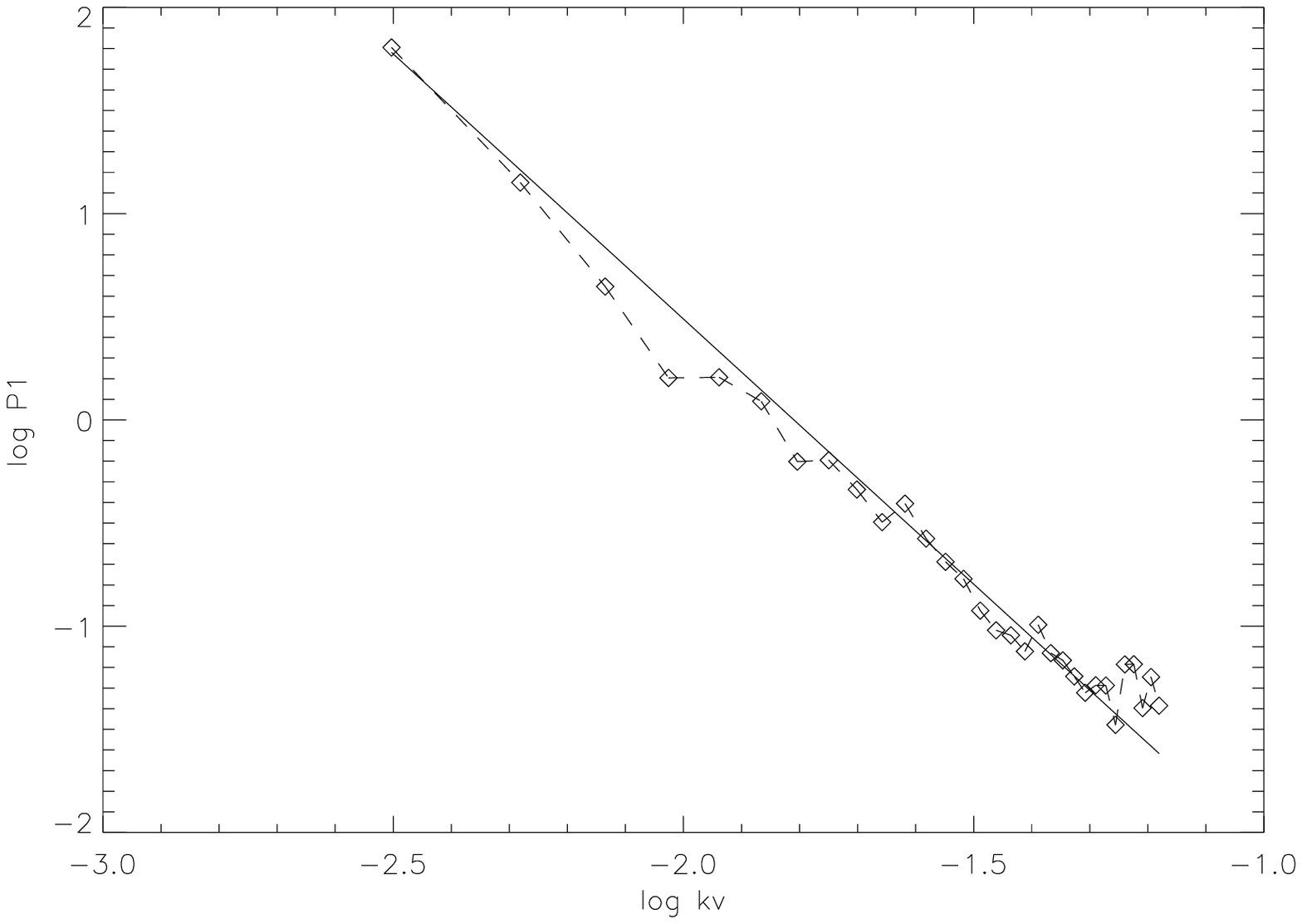} &
\inctab{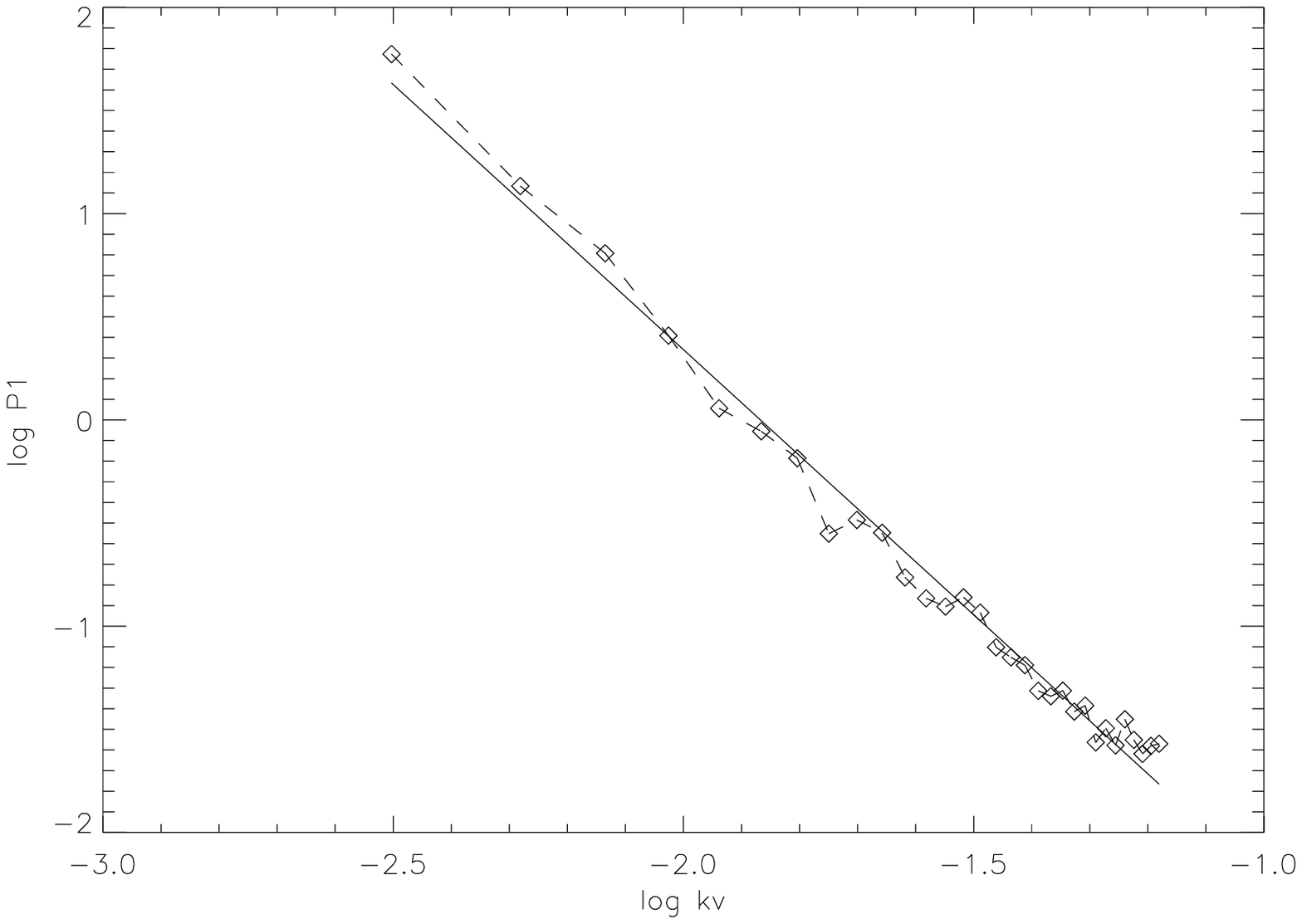} \\
\inctab{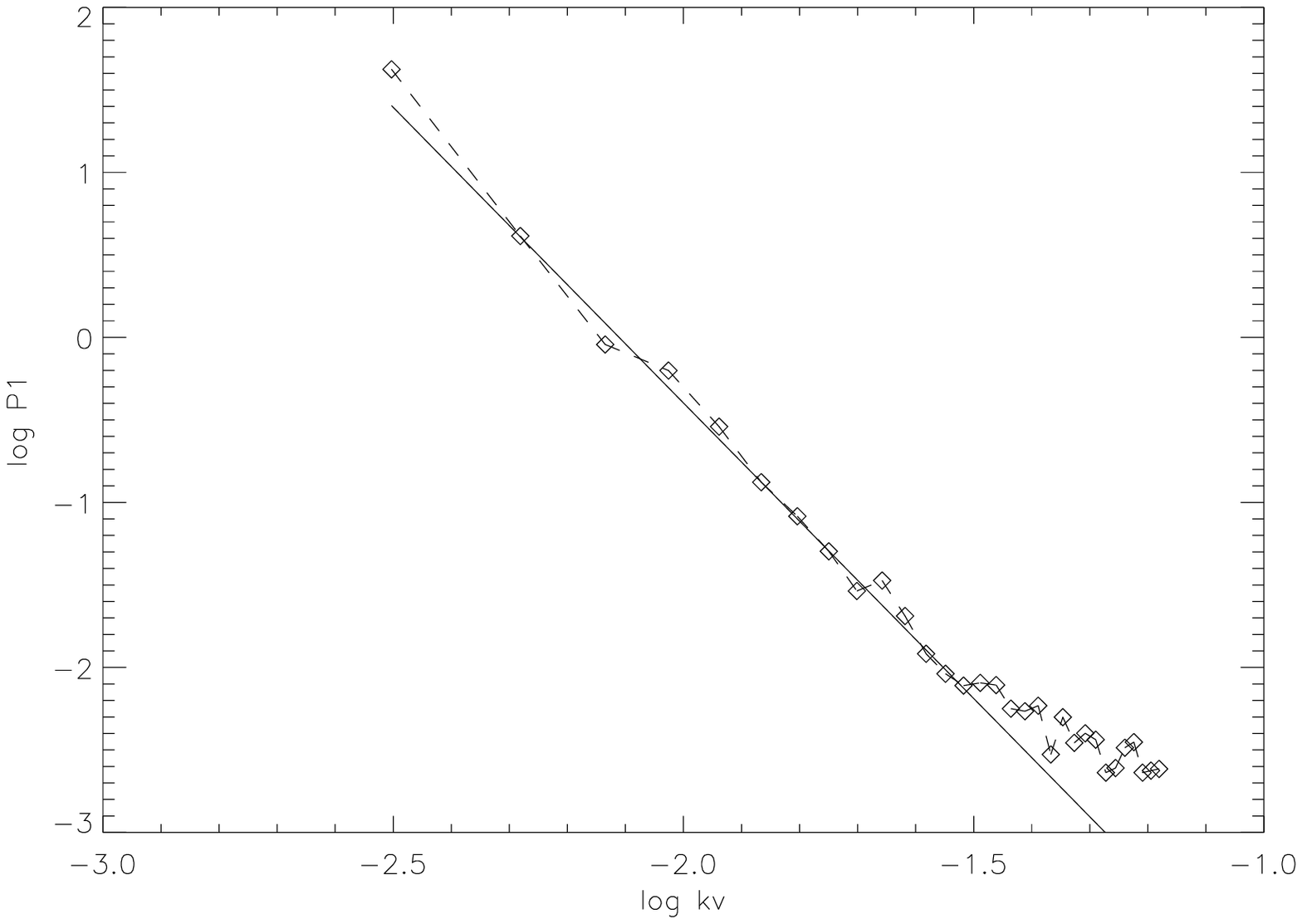} &
\inctab{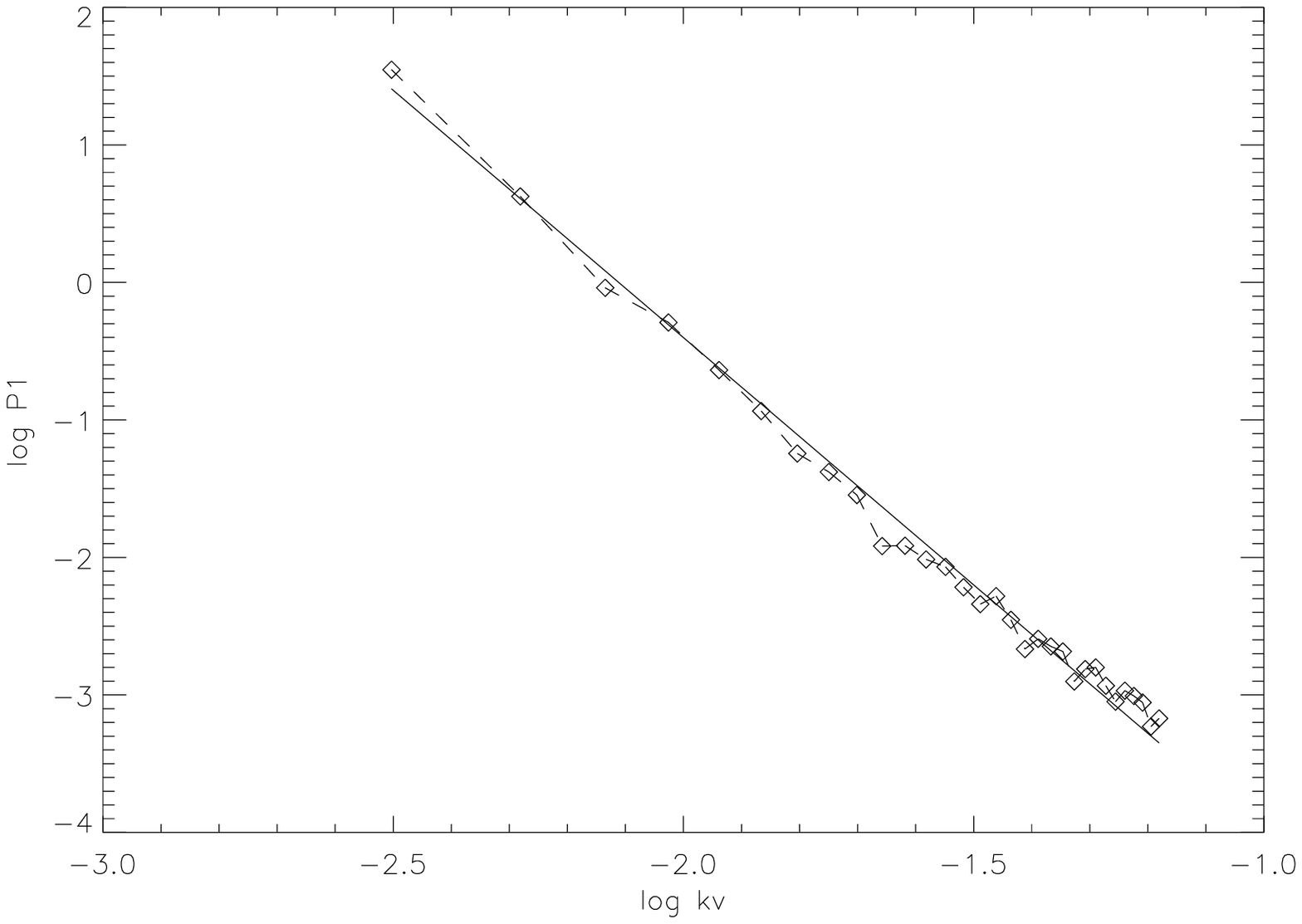} &
\inctab{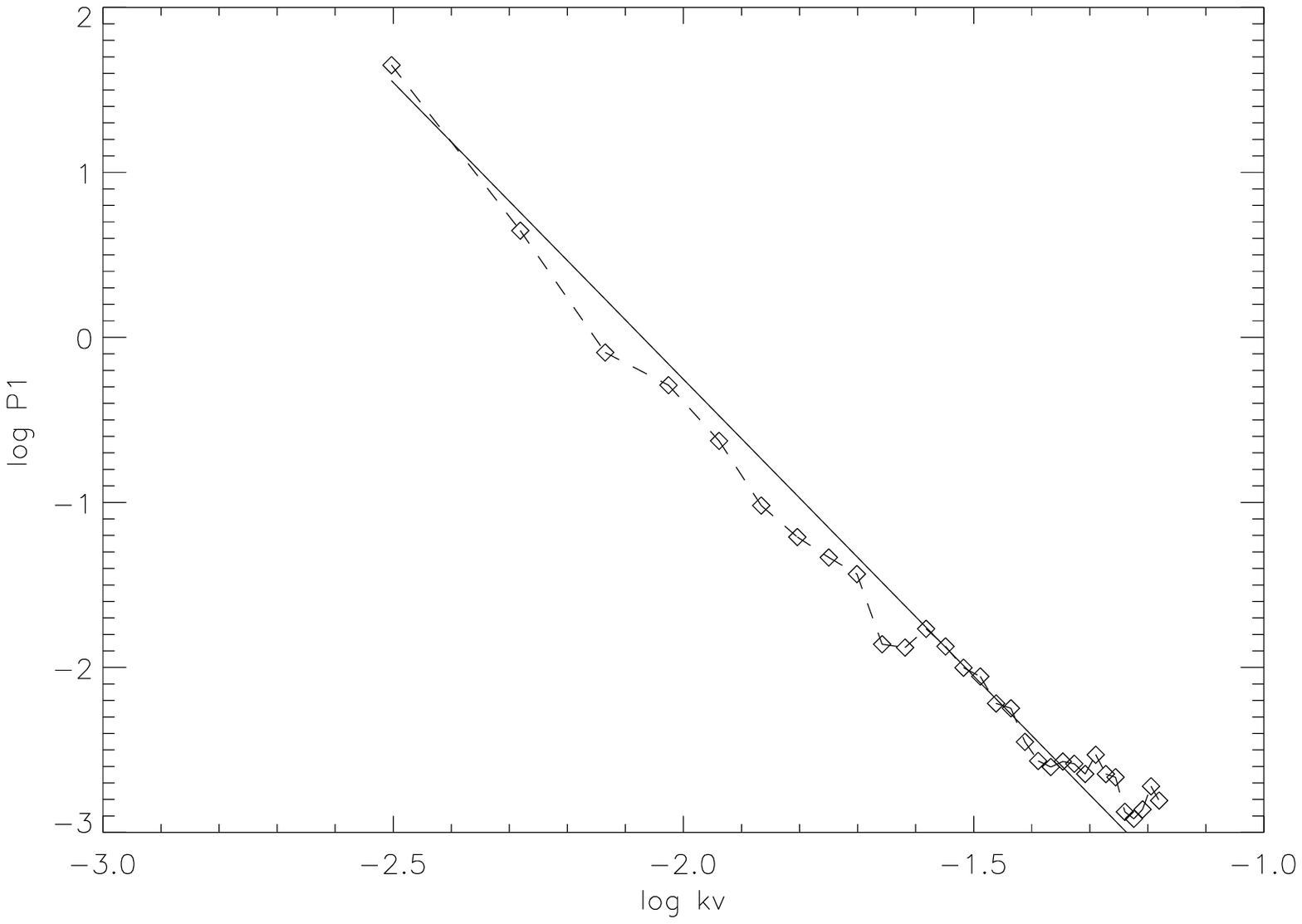}
\end{tabular}
\end{center}
\caption{Illustration to the requirement [\ref{eq:Nz}] to the number of points over line of sight. The rows correspond to different velocity spectral index $\alpha_v$ (4, 3.78 and 3.56 from top to bottom). VCS spectra in the middle column correspond to $N_z$ close to the estimation ($2^{14}$, $2^{17}$ and $2^{23}$ respectively), in the left column $N_z$ is 4 times lower, in the right -- 2 times greater. Undervalued $N_z$ increases shot-noise. (Solid lines show the expected slopes).\label{fig:Nz_req}} 
\end{figure}

VCS spectra, illustrating condition [\ref{eq:Nz}] are presented in Fig. \ref{fig:Nz_req}. The velocity spectral index $\alpha_v$ decreases from the top to the bottom (the values are 4, 3.78 and 3.56). According to our estimates above, this requires a drastic increase of number of points. Indeed, $N_z$ is, respectively, $2^{14}$, $2^{17}$ and $2^{23}$. In the left column with the shot-noise clearly present $N_z$ is 4 times undervalued, in the right one it is increased by factor two.

Examples of restoration of the velocity spectral indices from simulations are presented in Tab. \ref{tab:alpha_v_re}. They show good agreement, with the exception of somewhat outfalling value for $\alpha_v=4$. Values of $N_z$ have been chosen regarding Eq. [\ref{eq:Nz}].

\begin{figure}[p] 
\begin{center}
\plottwo{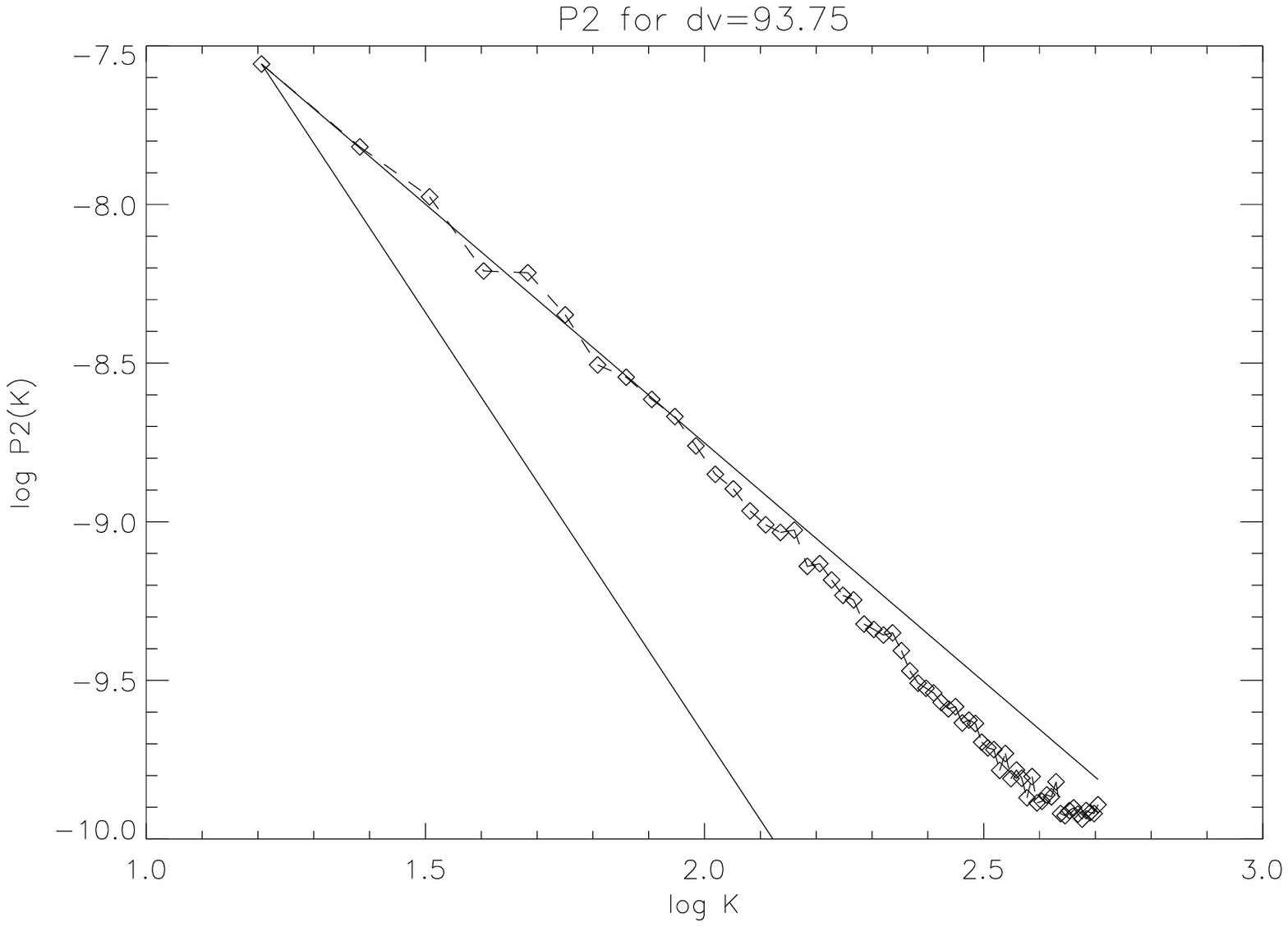}{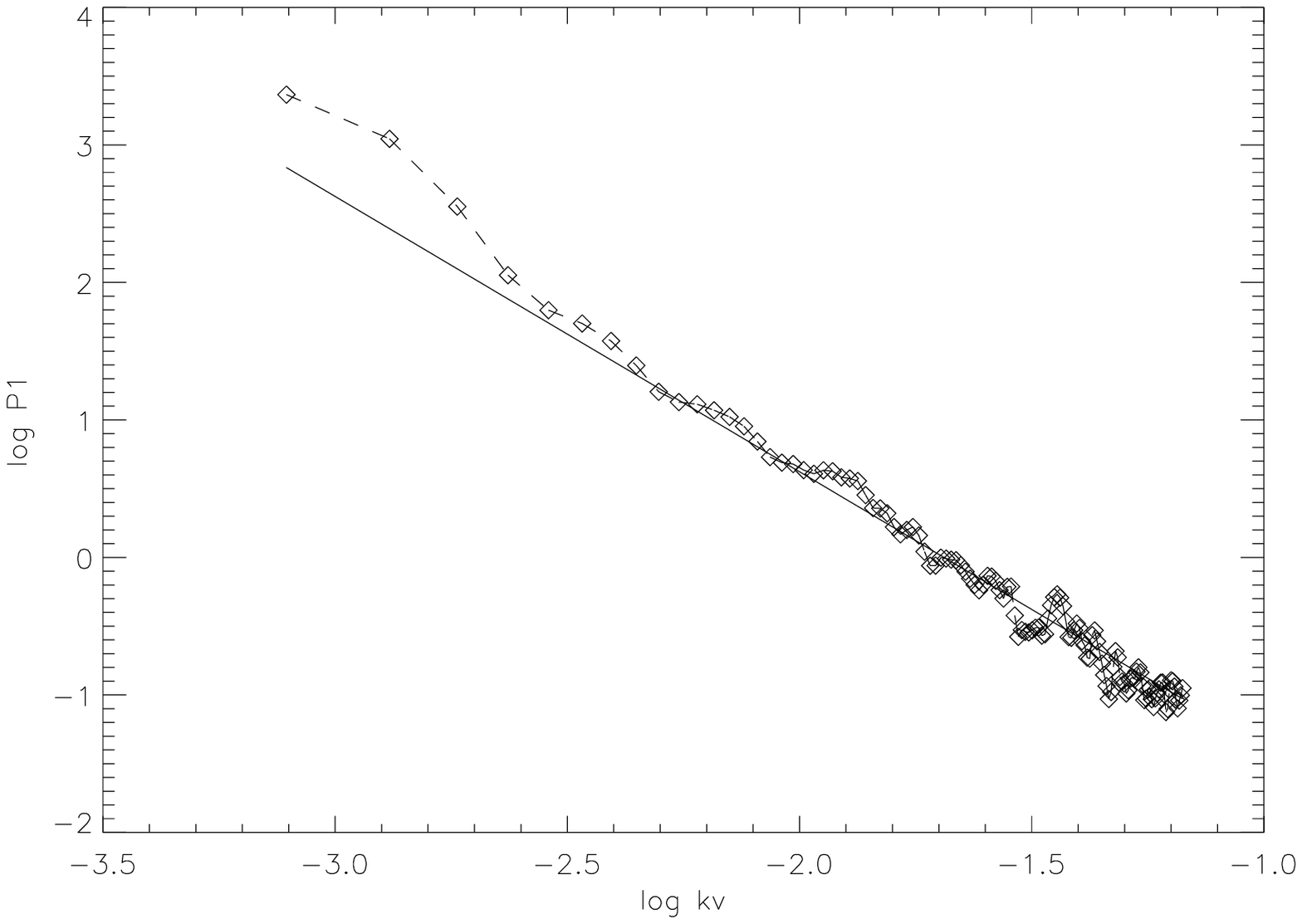}
\end{center}
\caption{Simulations with the number of points over line of sight $N_z=32768$, sufficient to suppress shot-noise ($N_z=20000$ is needed). Left: VCA spatial spectrum (2-d simulation), shallower solid line -- expected velocity-dominated spectrum, steeper solid line -- density-dominated spectrum. Right: correspondent VCS spectrum for high resolution mode (1-d simulation), solid line -- expected slope. Velocity spectral index is $4$. \label{fig:VCA_VCS_4_00}} 
\end{figure}

\begin{figure}[p] 
\begin{center}
\plottwo{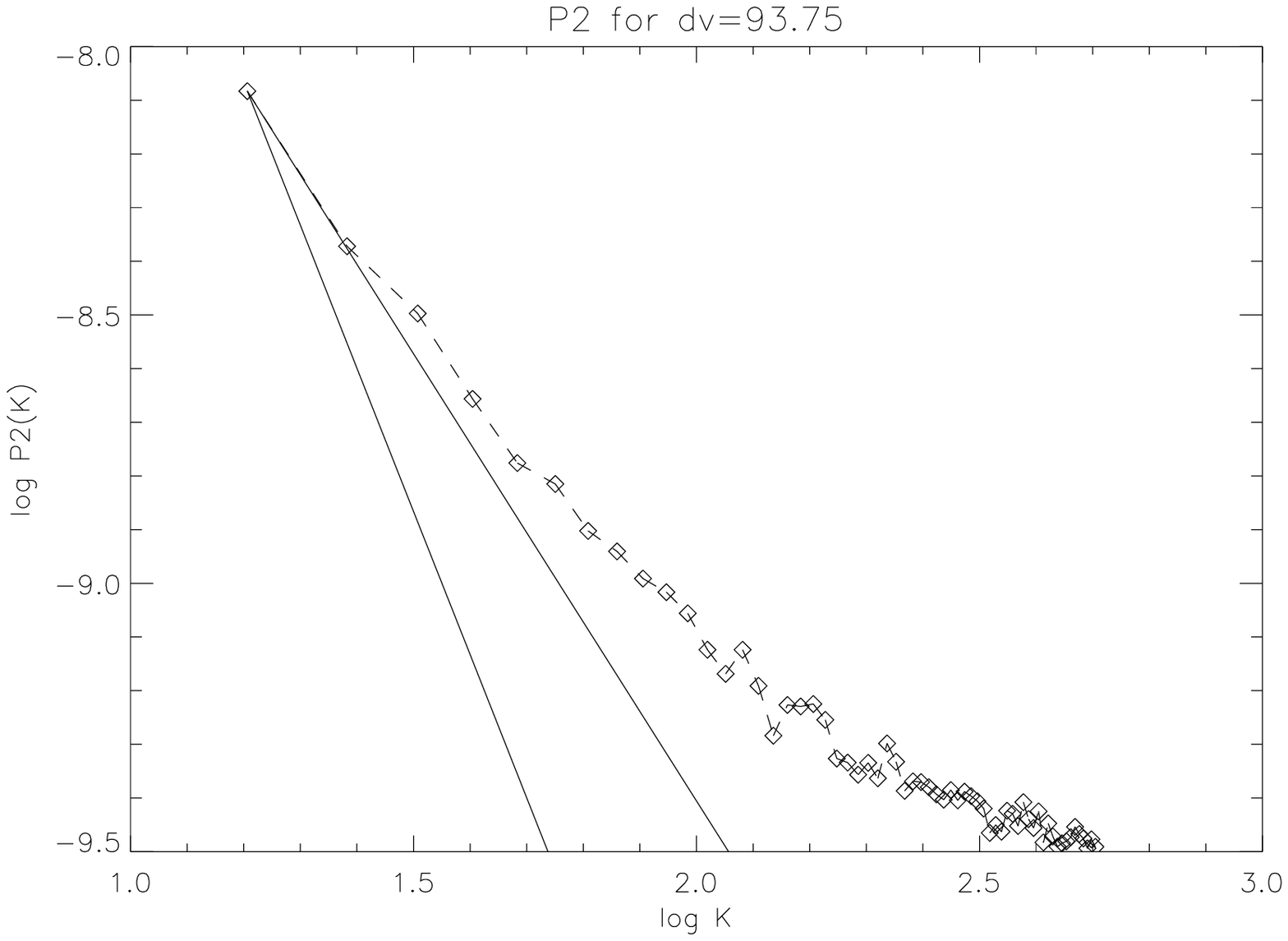}{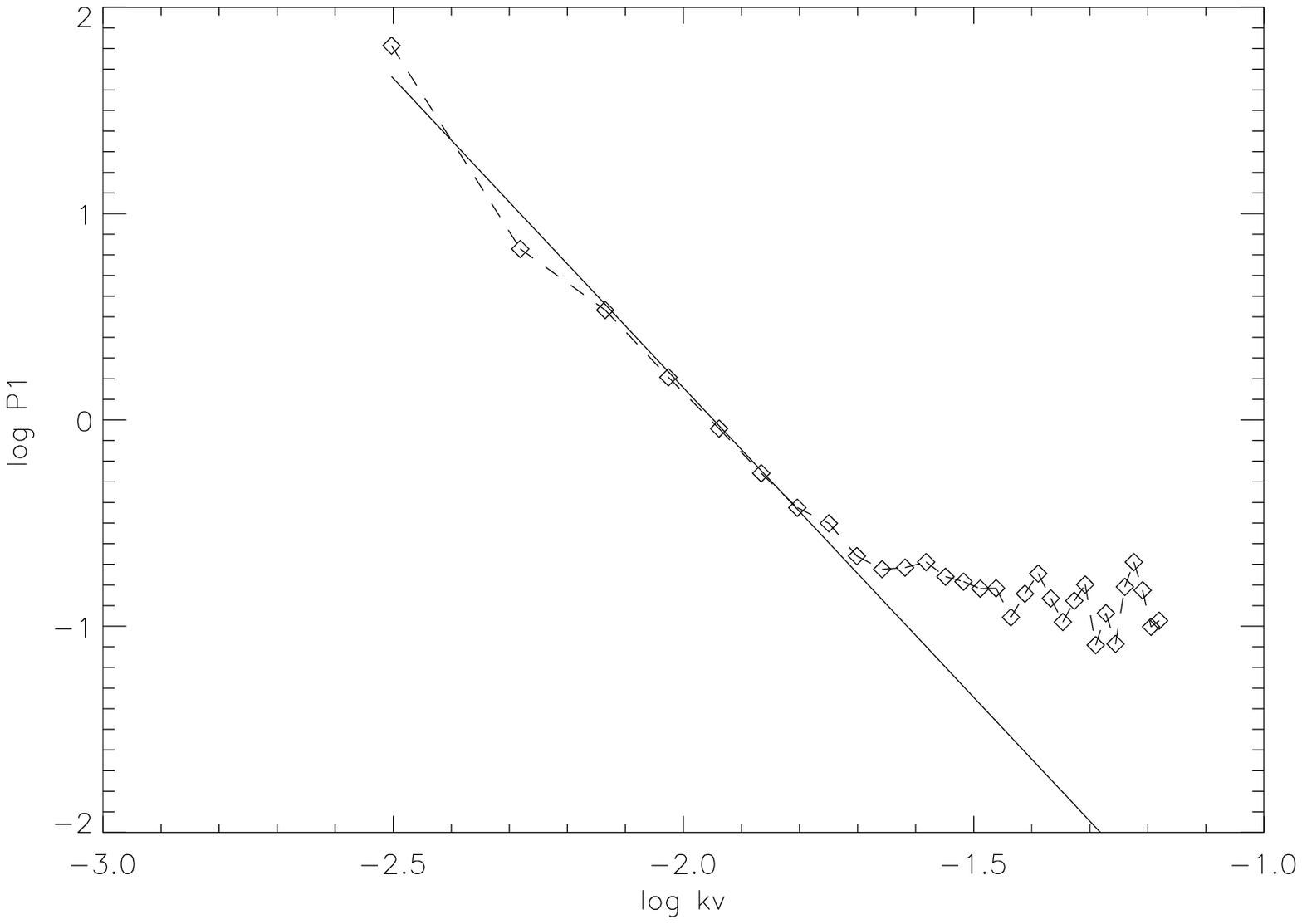}
\end{center}
\caption{Simulations with undervalued number of points over the line of sight ($N_z=32768$ instead of $N_z \approx 420000$). Left: VCA spatial spectrum (2-d simulation), shallower solid line -- expected velocity-dominated spectrum, steeper solid line -- expected density-dominated spectrum; the spectrum is distorted (the velocity-dominated one is expected). Right: correspondent VCS spectrum for high resolution mode (1-d simulation), solid line -- expected slope, shot-noise clearly visible. Velocity spectral index is $11/3$, which implies larger minimal $N_z$, than for the case on Fig. \ref{fig:VCA_VCS_4_00}.   \label{fig:VCA_VCS_3_67_unfilt}} 
\end{figure}

Knowing the requirements on $N_z$, we generate the VCS and VCA spectra (see Fig. \ref{fig:VCA_VCS_4_00}, with $\alpha_v=4$). But for shallower velocity spectrum satisfying such conditions for VCA may be difficult, even if we use 2-d data instead of 3-d data cube. For example, the condition [\ref{eq:Nz}] for the VCA simulations at $\alpha_v=11/3$ implies a square array with the side $420000$, which is difficult to implement at the moment. If we decrease this size, the VCA spectrum gets completely distorted, see Fig. \ref{fig:VCA_VCS_3_67_unfilt}, left. Therefore we have to seek alternative approaches. 

\begin{figure}[p] 
\begin{center}
\plottwo{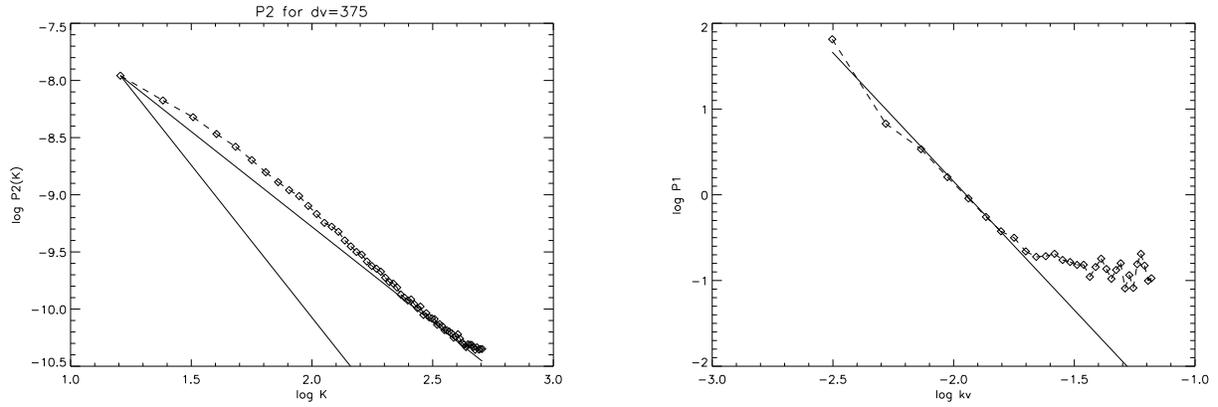}{vcs_3_67_1d_noisy.ps}
\end{center}
\caption{The same as Fig. \ref{fig:VCA_VCS_3_67_unfilt}, but with noisy $k_v$-harmonics filtered out when calculating the VCA spectrum (left). \label{fig:VCA_VCS_3_67_filt}} 
\end{figure}

In \cite{Esq03} it was shown, that if we use the thickest possible velocity slice, we get a correct spectrum. Let us consider, why this can happen. Looking at the right panel on Fig. \ref{fig:VCA_VCS_3_67_unfilt}, we see, that at higher wavenumbers $k_v$'s the VCS spectrum is affected by shot-noise. As both the VCA and the VCS spectra are just different projections of the same 3-d power spectrum in the PPV space (see Eqs. [\ref{eq:P2}] and [\ref{eq:P1}]), the noisy harmonics seen in the $P_1$ plot are most likely to affect the $P_2$ spectrum. If we use only $k_v$'s, which correspond to the intact part of $P_1$, the picture improves significantly, as it can be seen in Fig. \ref{fig:VCA_VCS_3_67_filt}, left. Such truncation of the $k_v$-domain is equivalent to taking a thicker velocity slice in \cite{Esq03}.

\begin{figure}[p] 
\begin{center}
\plottwo{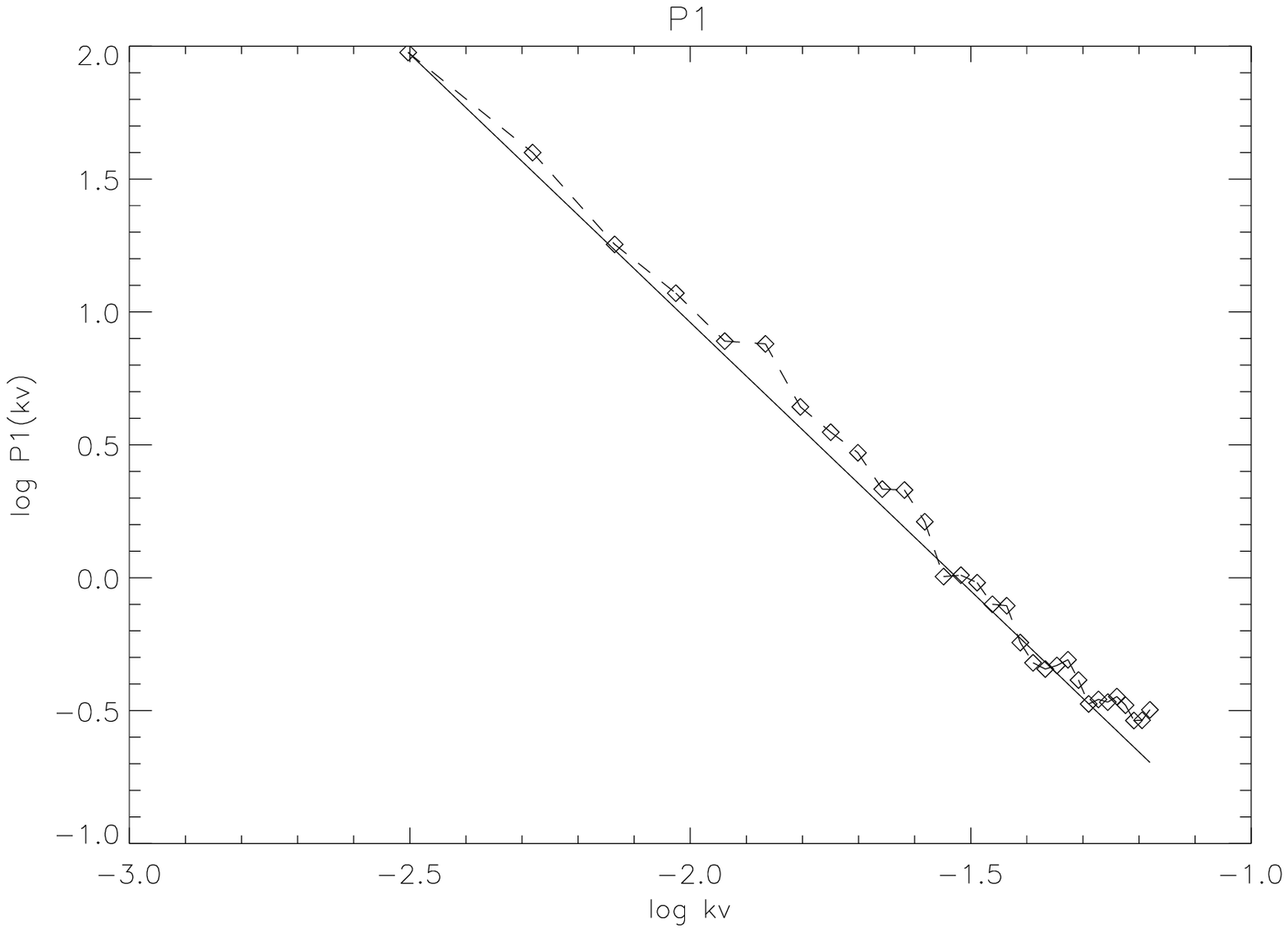}{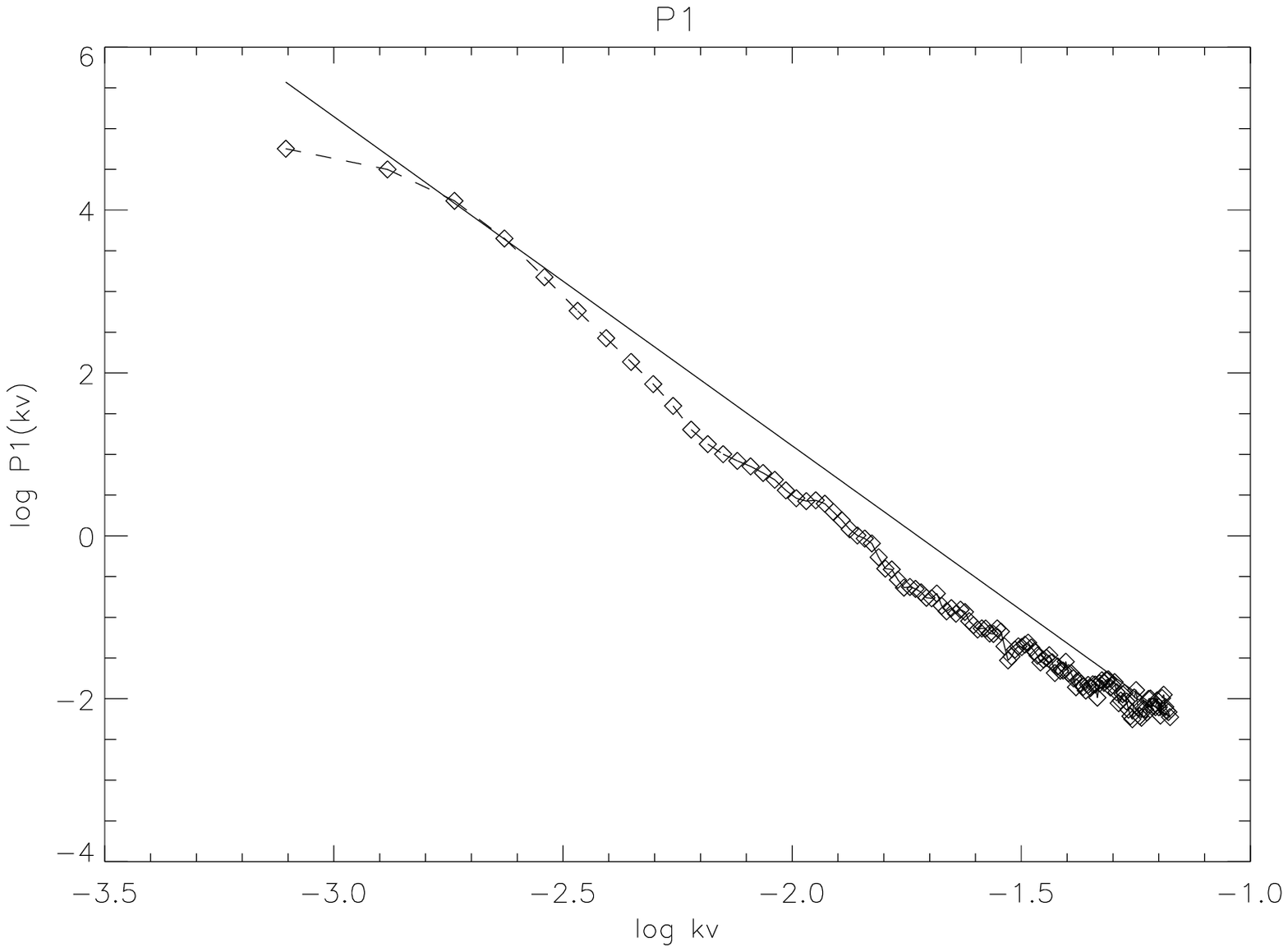}
\end{center}
\caption{2-d VCS simulations for high-resolution (left) and low-resolution (right) modes. Solid lines show expected slopes. Velocity spectral index is $4$. \label{fig:VCS_4_00_2d}} 
\end{figure}

It is also important to test the VCS prediction of steepening of $P_1$ when the beam is made wide enough to emulate the low resolution regime. To do so we used steeper velocity spectrum, with slope $4$, which is less demanding in terms of the number of points along line of sight. If we meet condition [\ref{eq:Nz}] we have a good agreement with the predictions for both high- and low-resolution regime, see Fig. \ref{fig:VCS_4_00_2d}.

Besides of testing of analytical results, synthetic data calculations could be helpful when pure analytical approach is difficult or impossible. For instance, we have tested if VCS technique is applicable in two-phase medium such as interstellar HI. It was speculated in LP00 that the fluctuations in the channel maps are dominated by cold (100~K) component of HI, while the warm (6000~K) component is subdominant for the VCA studies. This corresponds to the decrease of $\ft{f}$ in Eq. [\ref{eq:P1}].

In our simulations the phases separation was set by a threshold, applied to the density field. We assumed that the cold phase corresponds to the density larger than the threshold. We have also assumed that turbulence is the same for both phases, which can be justified from physical arguments (see \citealt{LP00}). 

\begin{figure}[p] 
\begin{center}
\includegraphics{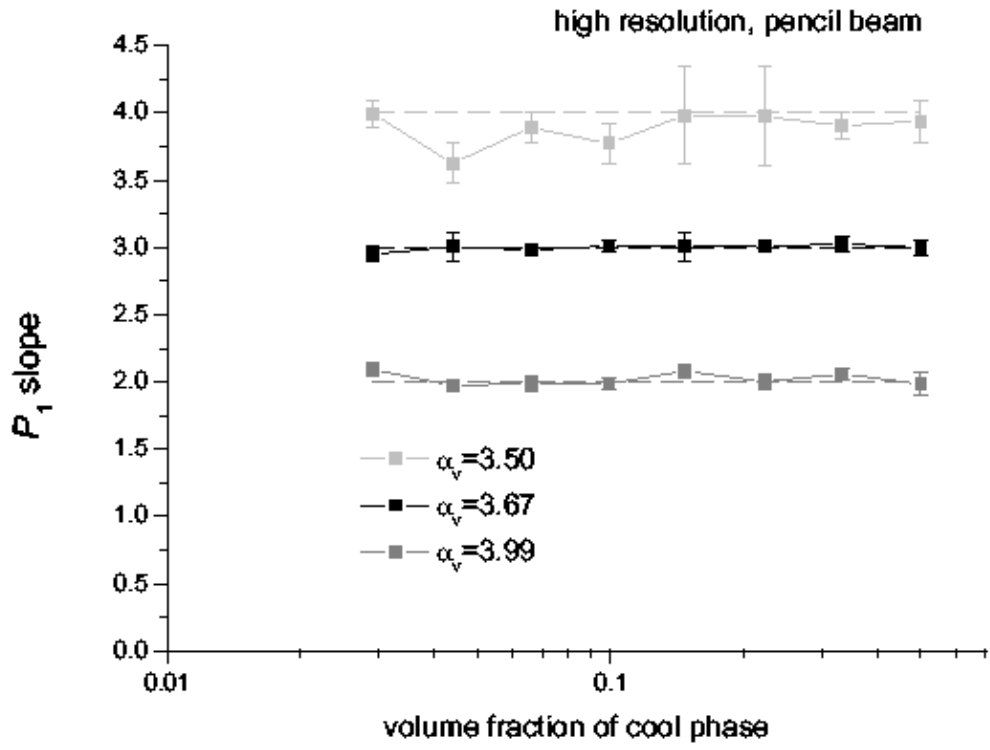}
\end{center}
\caption{Slope of $P_1$ as a function of an emitting medium volume (for different velocity spectral index $\alpha_v$). \label{fig:P1_slopes}} 
\end{figure}

In Fig. \ref{fig:P1_slopes} the slope of $P_1$ is plotted against the volume fraction of emitting medium for different spectral indices of velocity. As one can see, with our assumptions there is no visible influence of the fraction of the cold phase. We argue that the points for the shallow velocity spectrum have rather lower quality, because the condition of the shot-noise elimination [\ref{eq:Nz}] is not met in our simulations.

\section{Studies of absorption lines} \label{sect:sal}

\begin{figure}[p] 
\begin{center}
\plotone{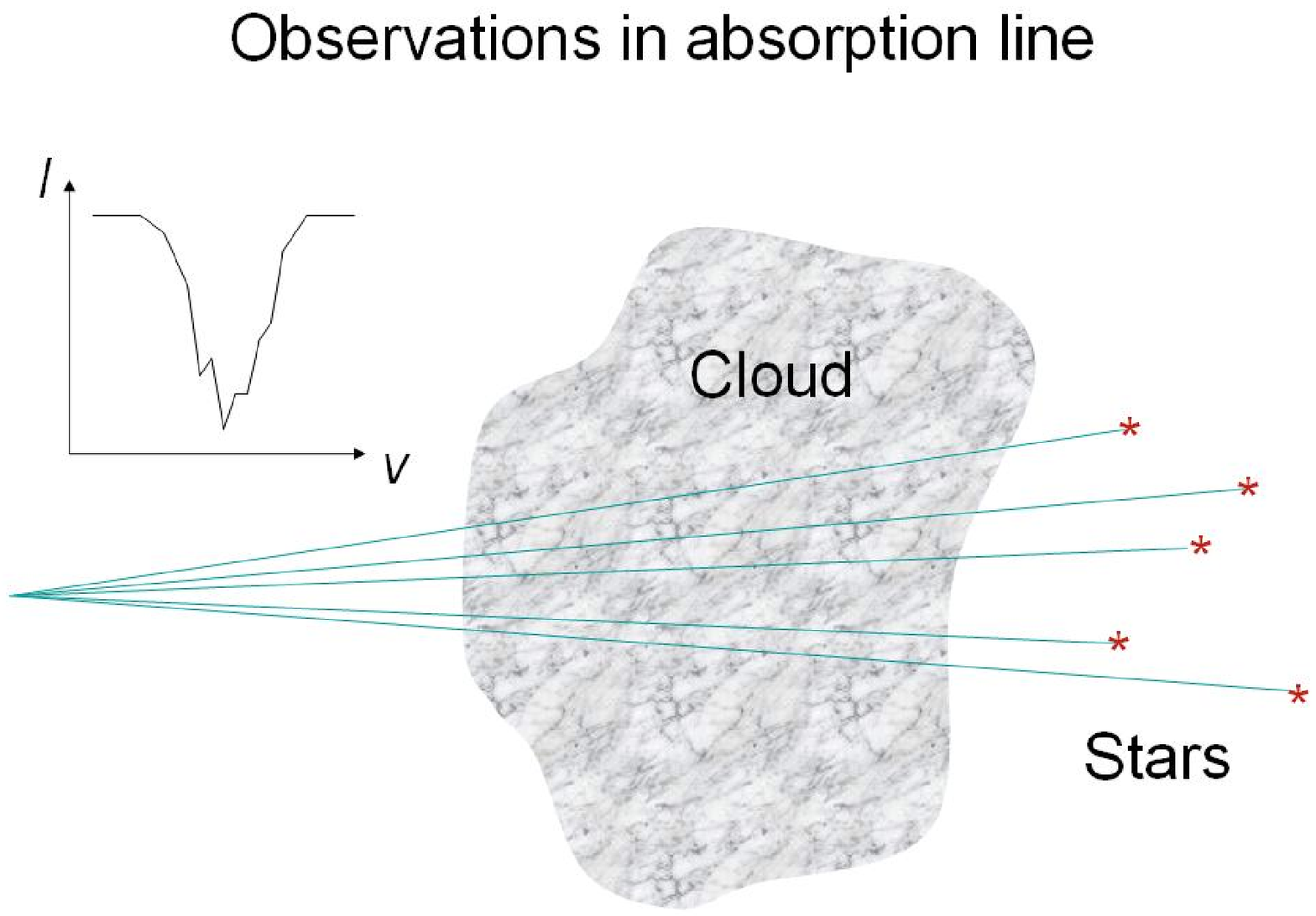}
\end{center}
\caption{VCS allows recovery of velocity statistics from absorption lines from stars. The whole $P_1$ is guaranteed to be in high-resolution mode, which provides better dynamical range over $k_v$. Numerical simulations show, that very few independent measurements are needed to gain required statistics.\label{fig:abs}} 
\end{figure}

\begin{figure}[p] 
\begin{center}
\plotone{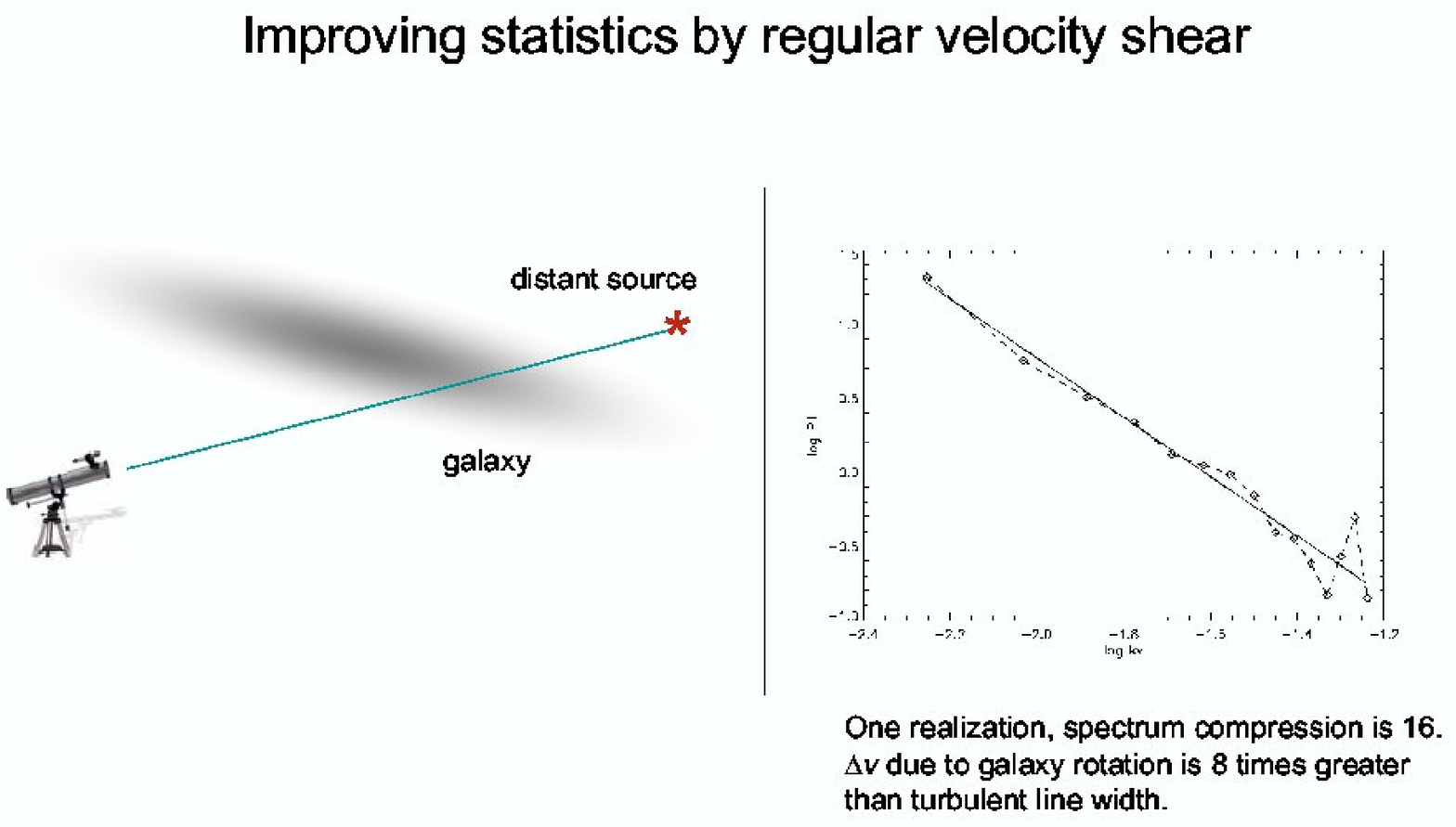}
\end{center}
\caption{VCS can be applied even to a single spectral line. Configuration can be as shown on this picture. Regular velocity shear from the galactic rotation results in statistics being sufficient to recover $P_1$\label{fig:shear}} 
\end{figure}

We must stress, that although fields of applicability of discussed techniques intersect, they do not coincide. It is quite evident, that VCA requires less resolution over velocity coordinate, and it can be used with lesser Mach number than the one needed for VCS. On the other hand, even if we have very poor statistics over angular coordinates and VCA doesn't work, VCS still can be used. It allows, for instance, working with absorption lines from distant sources, observed through the same turbulent medium, like shown on Fig. \ref{fig:abs}. The limiting case is using only one spectral line, but this requires significant regular velocity shear (it ``unfolds'' velocity fluctuations to wider spectral line and leads to better statistics, which we can utilize by averaging of adjacent $k_v$-harmonics, see Fig. \ref{fig:shear}). Without the shear, our simulations show, that only $~10$ independent measurements are needed to gain required statistics for $N_\sigma \approx 20$.


\section{Summary} \label{sect:sum}

We provided numerical testing of the predictions for the VCA and the VCS techniques, identified the source of shot noise reported in earlier testings, provided an analytical estimate of the resolution required to eliminate the noise and checked successfully this requirement.

As the VCA and VCS techniques are closely related (cf. Eqs. [\ref{eq:P2}] and [\ref{eq:P1}]), the presence of shot-noise in VCS inevitably affects the quality of VCA spectrum. One has to meet a rather demanding requirement given by Eq. [\ref{eq:Nz}] to eliminate the noise. 

Shot-noise appears in real observations when the number of received photons is low enough. Such could be X-ray or optical observations. Condition [\ref{eq:Nz}] gives in this case the required number of photons and can be used for estimation of observation time.

We found, that if we accurately deal with shot-noise, we get spectra, which are in agreement with predictions of VCA/VCS techniques. However the condition for shot-noise elimination can be very demanding for shallower velocity spectra ($N_z \to \infty$ exponentially when $\alpha_v \to 3$). This can make the 3-d and even 2-d simulations testing for the VCS and the VCA rather difficult.

We observed, that fragmentation of emitting medium due to presence of warm phase does not change the slope of $P_1$, predicted in VCS, which allows using this technique for neutral hydrogen in Milky Way in which both cold and warm gas are present.

Having numerical tools to test VCA/VCS techniques in various physical environments, we enhanced the reliability of the results obtained from observational data. 

\textbf{Acknowledgments} Authors thank Dmitry Pogosyan for his input. AC and AL acknowledge the NSF grant AST-0307869 and
AL thanks for the support from the Center for Magnetic Self-Organization in
Laboratory and Astrophysical Plasmas.


\appendix

\section{Coding}

The simulation programs were written in C++. There are many reasons for using this language. One of them is that we have a convenient infrastructure for managing of input data and initializing of an object hierarchy, provided by class \verb|data|. It gives us an easy way of setting up the input parameters without a necessity to recompile before each run. It also allows us to easily configure a batch job and hide details of MPI initialization, if the latter is used.

The input for the class \verb|data| is a set of text files with a usual ``id=value'' syntax, linked to each other in a tree. When we call the \verb|data| constructor with a name of a root file, this tree is mapped to a tree of objects of the type \verb|data|, each representing one input file.

The object structure of a simulation program is also a tree. Therefore it is convenient to map it to some subtree in the \verb|data| hierarchy. If we do this, constructor signatures rarely become more complex than \verb|MyClass::MyClass(data*)| and all data initialization is done by instantiating of a root object:

\begin{verbatim}
int main(void)
{
  data input("root.df") ;
  DProcessor dp(input.getnode("DPROCESSOR")) ;
  dp.GenerateAndProcess() ;
  dp.SaveResults() ;
  return 0 ; 
}
\end{verbatim}


Data files allow references defined by a path with respect to a root node. This allows us, for example, to avoid duplicating of identical data, or to put frequently changed data in one file. Such reorganizations do not require any changes in a program because references are transparent for client code.

This mechanism is also used for running a program in batch mode. In this case one file (``plugin'') contains all the information to be changed at each run and is re-generated by a controlling program before restarting of simulation. A separate node is created for defining this information, a ``parameter space''. Practically it means that we specify parameters to be scanned in a \verb|parspace.df| file and run our \verb|runtask| command, which takes care of \verb|plugin.df|, controls the execution and stores results of each simulation in a dynamically created directory tree.

\begin{figure}[p] 
\begin{center}
\includegraphics[scale=.35]{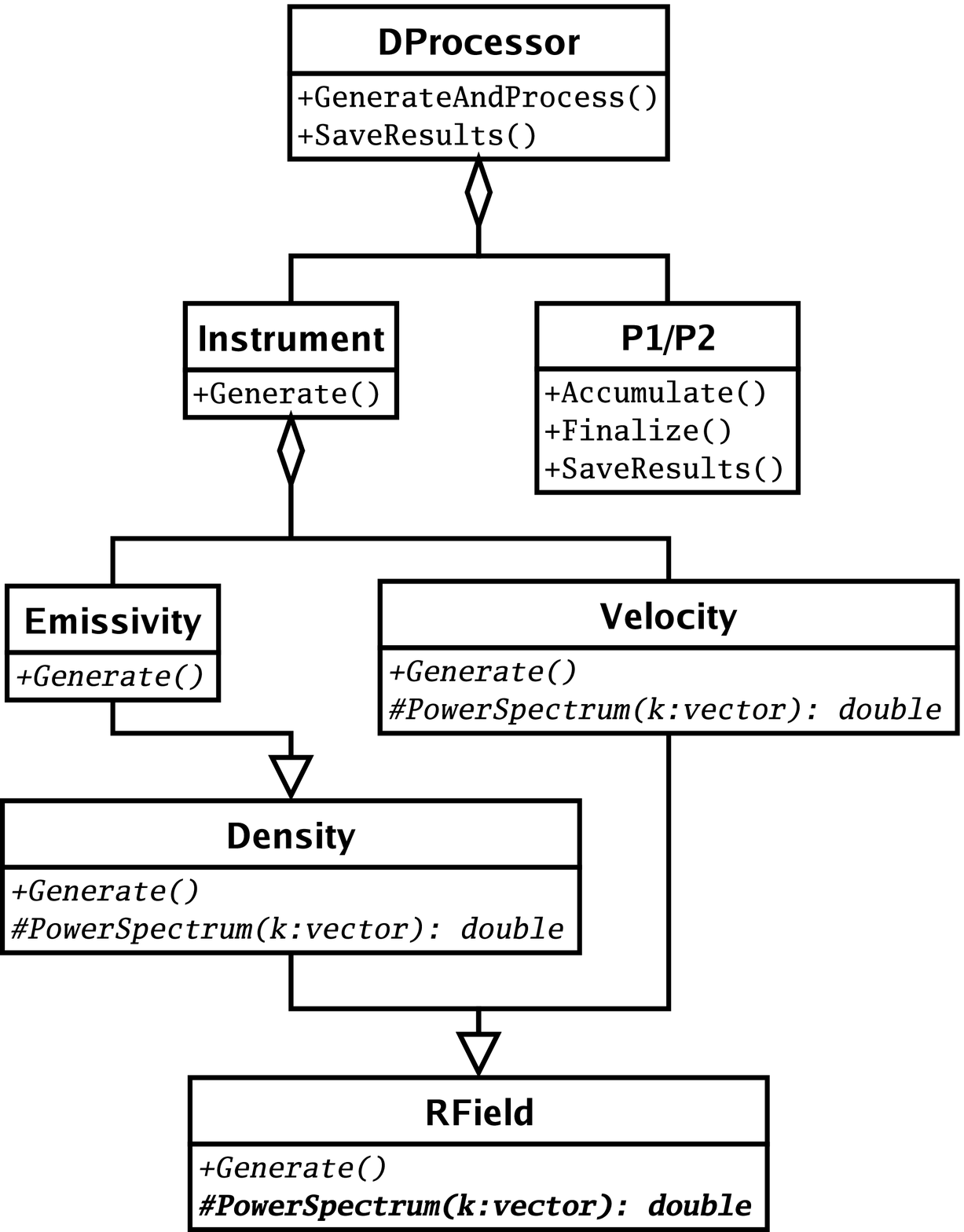}
\end{center}
\caption{Class diagram for VCS/VCA simulation programs. Class \texttt{RField} provides functionality for generating of random fields, class \texttt{Instrument} provides spectrometric data. Processing class is named \texttt{P1} for VCS and \texttt{P2} for VCA. Class \texttt{DProcessor} provides coordination between \texttt{Instrument} and \texttt{P1/P2}. This diagram is the same for versions with different number of dimensions and with or without parallelization. Triangle designates inheritance, rhomb -- aggregation. Polymorphic functions are given in italic.
\label{fig:classd}} 
\end{figure}

Class diagram for our programs, shown in Fig. \ref{fig:classd}, is approximately the same regardless of VCA or VCS version, number of dimensions, or presence of parallelization. Generation of random field is implemented in an abstract class \verb|RField|, from which classes \verb|Velocity| and \verb|Density| are derived. The function \verb|RField::PowerSpectrum(vector k)| is clean and needs definition in subclasses. This allows us to have different power spectra with the same random field generation code. Class \verb|Emissivity| implements emissivity law (linear or quadratic) and applies a spatial window function. Class \verb|Instrument| generates spectral data of required dimension and applies an instrument beam (in 2-d and 3-d VCS programs). These data are processed in class \verb|P1| (VCS) or in class \verb|P2| (VCA). Class \verb|DProcessor| controls generation and processing of realizations and accumulation of spectra.

The source of simulation programs is available at \sourcedir, see Tab. \ref{tab:progs} for the complete list. We also provide a utility for running simulations in batch mode, \verb|runtask|. The programs need installed IDL to produce graphics.

\newpage

\begin{deluxetable}{lll}
\tablecaption{Restoring of velocity spectral index $\alpha_v$ from simulations \label{tab:alpha_v_re}}
\tablehead{
  \colhead{expected $\alpha_v$} & 
  \colhead{restored $\alpha_v$} & 
  \colhead{Nz} 
}
\startdata
3.56 & 3.54$\pm$0.01 & 8388608 \\
3.67 & 3.65$\pm$0.02 & 524288 \\
3.78 & 3.79$\pm$0.02 & 131072 \\
3.89 & 3.89$\pm$0.02 & 32768 \\
4.00 & 3.91$\pm$0.04 & 16384 \\
\enddata
\tablecomments{16 realizations of spectral line have been used for each $P_1$}
\end{deluxetable}

\begin{deluxetable}{lll}
\tablecaption{VCA predictions about $P_2$ spectral index, velocity-dominated mode \label{tab:P2_slopes}}
\tablehead{
  \colhead{density spectrum} & 
  \colhead{2-d picture plane spectrum} & 
  \colhead{1-d picture plane spectrum}
}
\startdata
steep
&
\begin{math} 
  \frac{9-\alpha_v}{2}
\end{math}
&
\begin{math} 
  \frac{7-\alpha_v}{2} 
\end{math}
\\
shallow
&
\begin{math} 
  \frac{2\alpha_\eps-\alpha_v+3}{2}
\end{math}
&
\begin{math} 
  \frac{2\alpha_\eps-\alpha_v+1}{2}
\end{math}
\\
\enddata
\tablecomments{See \cite{LP00} for details}
\end{deluxetable}

\begin{deluxetable}{llll}
\tablecaption{VCS predictions about $P_1$ spectral index, parallel lines of sight \label{tab:P1_slopes}}
\tablehead{
  \colhead{density spectrum} & 
  \colhead{pencil beam} & 
  \colhead{flat beam} & 
  \colhead{low resolution}
}
\startdata
steep
&
\begin{math} 
  \frac{2}{\alpha_v-3}
\end{math}
&
\begin{math} 
  \frac{4}{\alpha_v-3} 
\end{math}
&
\begin{math}
  \frac{6}{\alpha_v-3}
\end{math}
\\
shallow
&
\begin{math} 
  \frac{2(\alpha_\eps-2)}{\alpha_v-3} 
\end{math}
&
\begin{math} 
  \frac{2(\alpha_\eps-1)}{\alpha_v-3}
\end{math}
&
\begin{math}
  \frac{2 \alpha_\eps}{\alpha_v-3}
\end{math}
\\
\enddata
\tablecomments{See \cite{LC06} for details}
\end{deluxetable}

\begin{deluxetable}{llll}
\tablecaption{List of simulation programs\label{tab:progs}}
\tablehead{
  \colhead{program} & 
  \colhead{dimensions} & 
  \colhead{techniques} & 
  \colhead{architecture}
}
\startdata
spect1d    & 1 & VCS & single processor\\ 
spect2d    & 2 & VCS & single processor\\ 
spect3d    & 3 & VCS & single processor\\ 
p\_spect2d & 2 & VCS & MPI\\ 
p\_spect3d & 3 & VCS & MPI\\ 
vca3d      & 3 & VCA & single processor\\ 
p\_vca2d   & 2 & VCA & MPI\\ 
p\_vca3d   & 3 & VCA & MPI\\
\enddata
\tablecomments{The programs are available at \protect{\sourcedir}}
\end{deluxetable}


\end{document}